\documentclass{article}
\usepackage[T1]{fontenc}
\usepackage[latin9]{inputenc}
\usepackage[letterpaper]{geometry}
\usepackage{textcomp}
\usepackage{amstext}
\usepackage{amsmath}
\usepackage{graphicx}
\usepackage{esint}
\usepackage[authoryear]{natbib}
\usepackage[nice]{nicefrac}

\makeatletter
\newcommand{\lyxaddress}[1]{
	\par {\raggedright #1
	\vspace{1.4em}
	\noindent\par}
}

\makeatother

\begin{document}
\title{\textbf{Evidence for multiple Ferrel-like cells on Jupiter}}
\author{Keren Duer$^{1*}$, Nimrod Gavriel$^{1*}$, Eli Galanti$^{1}$, Yohai
Kaspi$^{1}$,\\
 Leigh N. Fletcher$^{2}$, Tristan Guillot$^{3}$, Scott J. Bolton$^{4}$,
Steven M. Levin$^{5}$, \\
Sushil K. Atreya$^{6}$, Davide Grassi$^{7}$, Andrew P. Ingersoll$^{8}$,
Cheng Li$^{6}$, Liming Li$^{9}$,\\
 Jonathan I. Lunine$^{10}$, Glenn S. Orton$^{5}$, Fabiano A. Oyafuso$^{5}$,
J. Hunter Waite, Jr.$^{4}$}
\maketitle

\lyxaddress{\begin{center}
\textit{$^{1}$Department of Earth and Planetary Sciences, Weizmann
Institute of Science, Rehovot, Israel}\\
\textit{$^{2}$School of Physics and Astronomy, University of Leicester,
University Road, Leicester, LE1 7RH, UK}\\
\textit{$^{3}$Universitie Cote d'Azur, OCA, Lagrange CNRS, 06304
Nice, France}\\
\textit{$^{4}$Southwest Research Institute, San Antonio, Texas, TX,
USA}\\
\textit{$^{5}$Jet Propulsion Laboratory, California Institute of
Technology, 4800 Oak Grove Drive, Pasadena, CA 91109, USA}\\
\textit{$^{6}$Department of Climate and Space Sciences and Engineering,
University of Michigan, Ann Arbor, MI, USA}\\
\textit{$^{7}$Istituto di Astrofisica e Planetologia Spaziali, INAF,
Rome, Italy}\\
\textit{$^{8}$California Institute of Technology, Pasadena, California,
USA}\\
\textit{$^{9}$University of Houston, Houston, TX, USA}\\
\textit{$^{10}$Department of Astronomy, Cornell University, Ithaca,
New York 14853, USA}\\
\par\end{center}}

\lyxaddress{{*} These authors contributed equally to this work}

\subsubsection*{Key points}
\begin{itemize}
\item Measurements from multiple instruments of the Juno mission are interpreted
to reveal the meridional circulation beneath Jupiter's clouds
\item 16 Jet-paired deep cells, extending to at least 240 bar, are revealed
between latitudes $60^{\circ}{\rm S}$ and $60^{\circ}{\rm N}$, driven
by turbulence similar to Earth's Ferrel cells
\item The findings are supported by modeling the advection of tracers due
to the cells, showing agreement with ${\rm NH}{}_{3}$ data
\end{itemize}

\section*{Abstract}
Jupiter\textquoteright s atmosphere is dominated by multiple jet streams
which are strongly tied to its 3D atmospheric circulation. Lacking
a rigid bottom boundary, several models exist for how the meridional
circulation extends into the planetary interior. Here we show, collecting
evidence from multiple instruments of the Juno mission, the existence
of mid-latitudinal meridional circulation cells which are driven by
turbulence, similar to the Ferrel cells on Earth. Different than Earth,
which contains only one such cell in each hemisphere, the larger,
faster rotating Jupiter can incorporate multiple cells. The cells
form regions of upwelling and downwelling, which we show are clearly
evident in Juno\textquoteright s microwave data between latitudes
$60^{\circ}{\rm S}$ and $60^{\circ}{\rm N}$. The existence of these
cells is confirmed by reproducing the ammonia observations using a
simplistic model. This study solves a long-standing puzzle regarding
the nature of Jupiter\textquoteright s sub-cloud dynamics and provides
evidence for 8 cells in each Jovian hemisphere.

\newpage

\section*{Plain Language Summary}

The cloud layer of Jupiter is divided into dark and bright bands that
are shaped by strong east-west winds. Such winds
in planetary atmospheres are thought to be tied with a meridional
circulation. The Juno mission collected measurements of Jupiter's
atmosphere at various wavelengths, which penetrate the cloud cover.
Here we provide evidence, using the Juno data, of 8 deep Jovian circulation
cells in each hemisphere encompassing the east-west winds, gaining energy from atmospheric
waves, and extending at least to a depth of hundreds of kilometers.
Different than Earth, which has only 1 analogous cell in each hemisphere, known as a Ferrel cell, Jupiter can
contain more cells due to its larger size and faster spin. To support
the presented evidence, we modeled how ammonia gas would spread under
the influence of such cells and compared it to the Juno measurements.
The presented results shed light on the unseen flow structure beneath
Jupiter's clouds.

\section{Introduction}

Over the last few decades, spacecraft and ground-based observations
have gathered data about Jupiter's atmosphere, including measurements
of cloud reflectance \citep{garcia2001}, winds \citep{Porco2003,Salyk2006,Tollefson2017},
composition \citep{taylor2004} and lightning flashes \citep{little1999}.
Since 2016, the Juno spacecraft has provided unprecedented measurements
that revealed new information on the deep dynamics of Jupiter \citep{Bolton2017}.
Gravity science enabled an accurate mapping of Jupiter's gravitational
field \citep{Iess2018}, resulting in the inference that the zonal
jets penetrate $\sim3000$~km deep \citep{Kaspi2018,Guillot2018},
{where they possibly decay due to magnetic drag \citep{Liu2008,Dietrich2018,kaspi2020} and may also require the presence of a stable layer \citep{christensen2020}}.
The Jovian Infrared Auroral Mapper (JIRAM) provided measurements of
tropospheric species distribution below the cloud level \citep{grassi2020}.
The Microwave Radiometer (MWR) measurements, inferred as brightness
temperature ($T_{{\rm b}}$), revealed the deep ammonia abundance
\citep{Li2017,oyafuso2020}, as well as lightning at a frequency of
600~MHz \citep{brown2018}. The combination of these observations
allows the essential nature of Jupiter's deep overturning circulation
to be revealed, as the flows associated with such circulation are
directly related to cloud formation, temperature variations, lightning
occurrences, tracer distributions and turbulence.

Earth's atmosphere is commonly referred to as possessing a three-cell
meridional structure in each hemisphere \citep{Vallis2006}, which
can be recognized in the zonal-averaged velocities. Circulation cells
of such nature are thought to prevail in the atmospheres of terrestrial
planets \citep{read2018} and were observed, for example, on Mars
\citep{lewis2007} and Venus \citep{limaye2007}. On the terrestrial
planets, the solid surface drag plays a part in maintaining the circulation
in the cells. However, as the giant planets hold no such surface,
the mere possibility of them possessing meridional circulation cells
remained uncertain. Earth's midlatitudes are governed by the Ferrel
cells, which are driven by atmospheric turbulence, creating regions of eddy momentum flux convergence at midlatitudes \citep{Vallis2006}.
These cells accompany the midlatitude jets and are connected to the
cloud structure in Earth's atmosphere.

\begin{figure}
\begin{centering}
\includegraphics[width=0.75\textwidth]{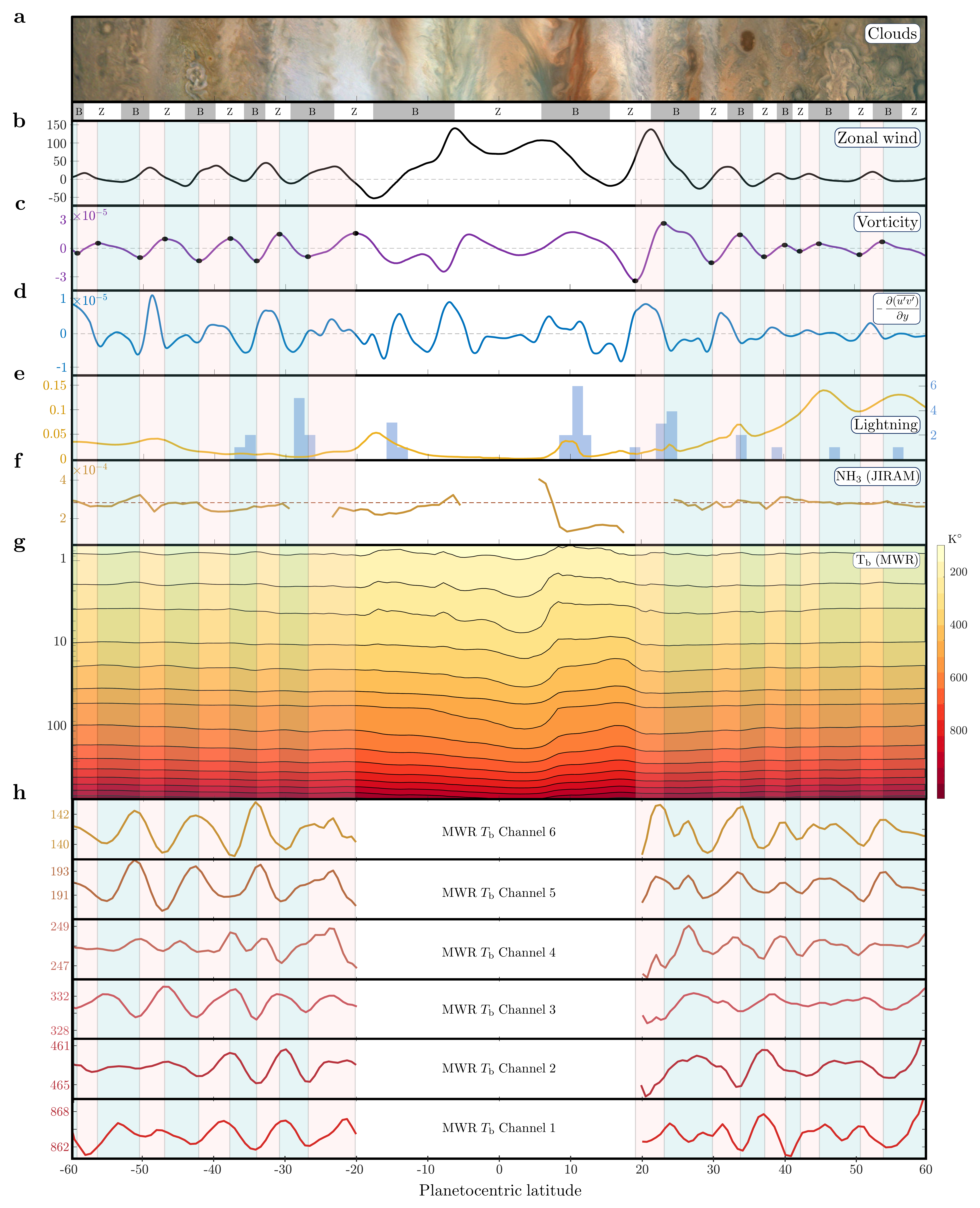}
\par\end{centering}
\caption{\label{fig:intro} Observations of Jupiter's atmosphere. (a) Image
of Jupiter's clouds (longitudes $69-87^{\circ}$) taken by JunoCam
on Dec.~26th~2019 during perijove 24 (image credit: NASA/JPL/SwRI/MSSS/Gerald~Eichstaedt/John~Rogers),
with the traditional ``dark'' belts (``bright'' zones) defined
as regions of cyclonic (anticyclonic) vorticity, identified below
as 'B' ('Z'). (b) {Jupiter's zonally averaged zonal
wind {[}${ \pm 15\;{\rm m\,\,s^{-1}}}${]} measured by the Hubble space telescope
on December 11, 2016, during Juno's third perijove \citep{Tollefson2017}.}
(c) The zonally averaged vorticity $[{\rm s}^{-1}]$, calculated from
the zonal wind profile (panel b). Black dots represent local extrema
in the midlatitudes. (d) { Eddy momentum flux convergence {[}${\pm 2\times10^{-6}\;\rm m\,\,s^{-2}}${]}
calculated from 58 image pairs taken by Cassini during its Jupiter
flyby in December, 2000 \citep{Salyk2006}}. (e) Lightning detections
$[{\rm s}^{-1}]$ by Juno's MWR during perijoves 1-8 \citep[yellow, left axis, ][]{brown2018} and number of lightning storms detected by the Cassini during its flyby \citep[blue, right axis, ][]{Porco2003}.
(f) Distribution of ammonia {[}volume mixing ratio{]} and its mean
(dashed) at a depth of $\sim6$ bar, measured by Juno's JIRAM during
perijoves 1-15 \citep{grassi2020}. (g) Nadir $T_{{\rm b}}$
{[}$^{\circ}{\rm K}${]} (color) interpolated between pressure levels
of $0.7$ and $240$~bar (vertical axis), measured by Juno's MWR
during perijoves 1-12 \citep{oyafuso2020}. {(h) Reconstructed MWR Brightness temperature at midlatitudes. A frequency filter is applied according to Eq.~S7. The standard deviation of each channel and latitude is available in Fig.~S3 and Fig.~S4. It can be seen that $T_{{\rm b}}$ changes its trend at the borders between cells, consistent with the Ferrel-like cells hypothesis.} (b-h) Light red (blue) bands in the midlatitudes indicate regions of positive (negative) vorticity
gradients.}
\end{figure}

The prominent banded structure at the cloud tops of Jupiter's atmosphere
(Fig.\ \ref{fig:intro}a) has been observed for centuries \citep{Vasavada2005}.
These reflectivity contrasts are partially aligned (mainly at low
latitudes, \citet{Ingersoll2000}) with the belts and zones (Fig.\ \ref{fig:intro}a),
defined by the sign of the zonal-wind vorticity ($\bar{\zeta}=-\nicefrac{\partial\bar{u}}{\partial y}$
in Fig.\ \ref{fig:intro}c, where $u$ is zonal velocity, $y$ is
in the meridional direction and an over-line represents a zonal mean).
Voyager measurements suggested that the zones are associated with
ascending motion, but this was limited to low latitudes due to its
equatorial trajectory and to altitudes above $0.5$~bar \citep{Gierasch1986}.
The latitudinal profile of the zonal wind, calculated using cloud-tracking
\citep{garcia2001,Porco2003,Tollefson2017}, reveals that the equatorial
region is characterized by a strong eastward flow, while the midlatitudes
exhibit alternating jets, spaced $2-8^{\circ}$ apart in latitude
(Fig.~\ref{fig:intro}b). The midlatitude jets are correlated with
the eddy momentum flux convergence \citep{Salyk2006} ($\nicefrac{-\partial\left(\overline{u'v'}\right)}{\partial y}$
in Fig.~\ref{fig:intro}d, where $v$ is meridional velocity and
an apostrophe represents deviations from the zonal mean, i.e. ``eddy''
terms), implying that the midlatitude jets are eddy-driven \citep{Ingersoll2000,young2017},
similar to the jets within Earth's Ferrel cells \citep{Schneider2006a,Vallis2006}.
To illustrate the relation between the jets and the eddies, regions
of positive (negative) vorticity gradient, $\nicefrac{\partial\zeta}{\partial y}=-\nicefrac{\partial^{2}\bar{u}}{\partial y^{2}}$,
at midlatitudes, are marked by light red (blue) bands (Fig.~\ref{fig:intro}),
where counter-clockwise (clockwise) Ferrel-like circulation cells
are expected in the northern hemisphere (NH). Similar circulations,
but in opposite directions, apply for the southern hemisphere (SH).
Evidence for vertical motion comes also from observations of lightning
flashes \citep{little1999,Porco2003,brown2018}, suggesting updrafts
in cyclonic belt regions (e.g., Fig.~\ref{fig:intro}e).

Additional information regarding Jupiter's deep atmosphere can be
obtained by probing Jupiter's interior at microwave frequencies. Juno's
MWR has 6 microwave channels \citep{janssen2017}, each measuring
the atmospheric $T_{{\rm b}}$ at a different depth \citep{janssen2017,Bolton2017,oyafuso2020,Fletcher2021},
and collectively covering the range between $\sim0.7$ and $\sim240$~bar
(Fig.~\ref{fig:intro}g,h, see also supporting information - SI).
$T_{{\rm b}}$ measurements are affected by both ammonia abundance and
temperature [and water in the case of the longest wavelengths, \citet{Li2017,li2020,Fletcher2021}].
If the latitudinal gradients of $T_{{\rm b}}$ were primarily driven
by temperature changes, then thermal wind balance implies that the
midlatitude jets strengthen from the cloud-deck to about $\sim8$~bar,
and then decay slowly towards the interior \citep{Fletcher2021}.
However, interpreting $T_{{\rm b}}$ as temperature would also imply
that the equatorial wind double its magnitude below the cloud level
\citep{Bolton2017}, which is inconsistent with gravity constraints \citep{duer2020}.
{Thus, the latitudinal variation of $T_{{\rm b}}$
is probably governed by ammonia opacity, resulting in a map of ammonia abundance
\citep{Li2017}, and implying that the zonal winds are nearly barotropic
\citep{Fletcher2021}}. The overall ammonia structure, supported also
by earlier observations \citep{DePater2001}, reveals stratification
of ammonia with depth, although the mean ammonia profile changes the
sign of its vertical gradient at the $\sim2-8$ bar region \citep{giles2017,Li2017,dePater2019}.
The atmospheric depletion and stratification of ammonia is likely
linked to small-scale storm activity \citep{guillot2020b,guillot2020a},
where water-ammonia hail, forming around the 1-bar level, falls below
the water-cloud base and releases ammonia and water at altitudes below
10~bar \citep{guillot2020a}. Additional measurements of ammonia
come from Juno's JIRAM, which evaluated the ammonia distribution at
a depth of $\sim5-6$~bar \citep{grassi2010,grassi2020} (Fig.~\ref{fig:intro}f),
indicating, as the MWR measurements, that ammonia varies with latitude.
These variations are the key observation for this study, as ammonia
anomalies (deviations from the isobaric mean) can reveal details about
Jupiter's overturning circulation \citep{ingersoll2017b,Fletcher2021,Lee2021}.

\section{Ammonia anomalies due to vertical advection\label{subsec:Ammonia-anomalies-due}}

\begin{figure}
\begin{centering}
\includegraphics[width=0.8\textwidth]{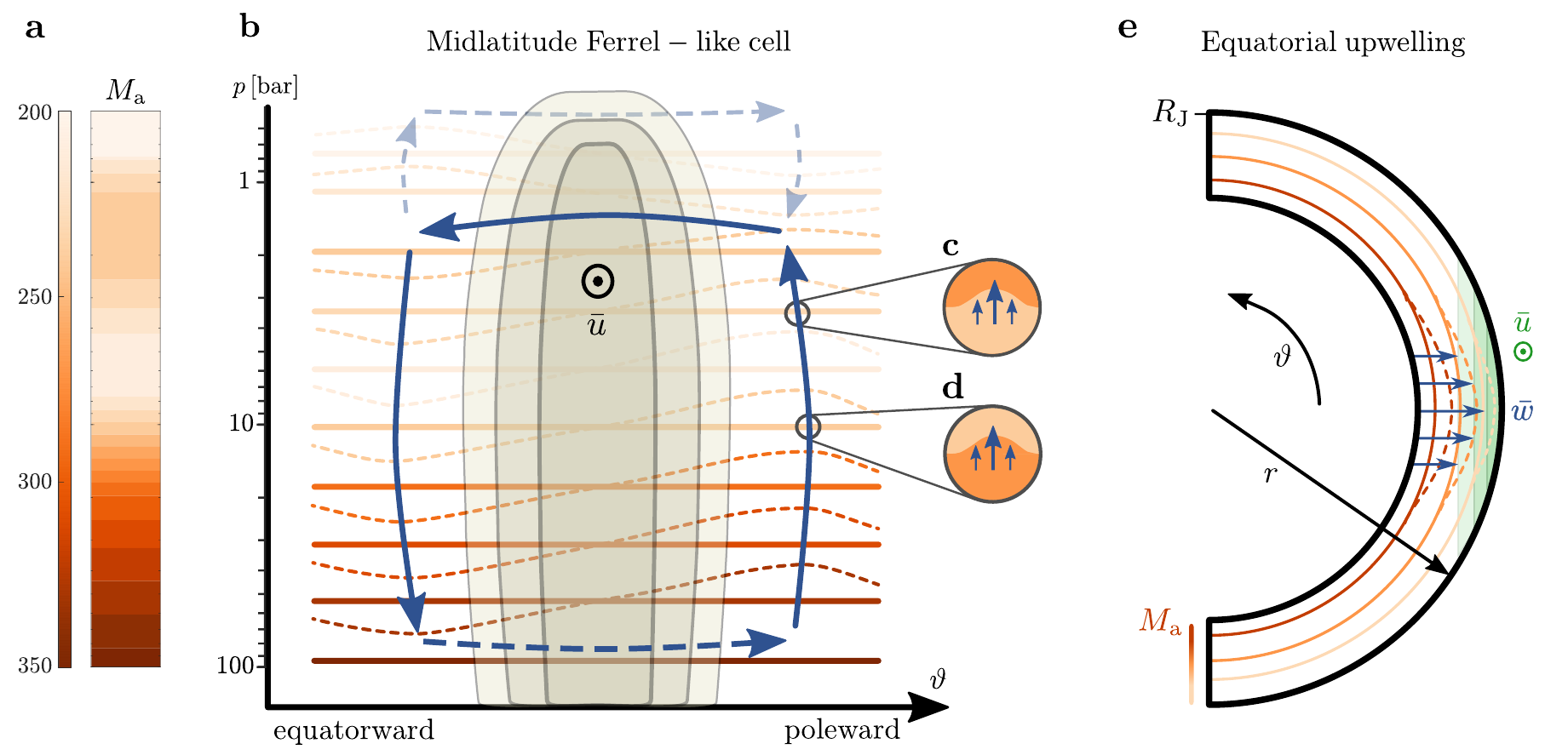}
\par\end{centering}
\caption{\label{fig:theory} Schematics of Jupiter's meridional circulation
as inferred from the ammonia distribution. (a) The vertical structure
of the meridionally averaged ammonia concentration ($M_{{\rm a}}$)
{[}ppm{]}, as interpreted from the $T_{{\rm b}}$ data \citep{Li2017}.
(b) Illustration of a midlatitude Ferrel-like circulation cell (blue
arrows) in the NH, looking from east towards the west. The cells are
accompanied by an eddy-driven barotropic jet ($\bar{u}$), which peaks
at the center of the cell (beige contours). Ammonia constant-concentration
lines are illustrated with orange shades (according to the ammonia
vertical profile from panel a). Dashed orange lines are deviations
from $M_{{\rm a}}$, driven by vertical advection. The return flow
of the cell, illustrated by a dashed blue arrow, lies at an unknown
depth. An oppositely directed upper cell, as suggested by pre-Juno
measurements \citep{Ingersoll2000,Showman2005,fletcher2020}, is demonstrated
by dashed transparent arrows. $p$ is pressure, taken as a vertical
coordinate. (c) A closer look at the region where the rising
air advects ammonia-poor fluid to an ammonia-rich layer, associated
with pressure levels between $1.5$ and $6$ bar. (d) Here, rising
gas drags higher ammonia concentration to a lower ammonia concentration
region, associated with pressure levels deeper than $6$ bar. (e)
A cross section of Jupiter's equatorial upwelling ($\overline{w}$),
associated with a superrotating jet ($\bar{u}$, green contours),
leading to ammonia concentration maximum. The equatorial $M_{{\rm a}}$ (orange contours) is assumed to decrease with radius (Fig.~S8).
$R_{{\rm J}}$ is Jupiter's radius and $\vartheta$ and $r$ are the
latitudinal and radial directions, respectively.}
\end{figure}

In the presence of a stable vertical ammonia concentration gradient,
advection by the vertical branches of a meridional circulation can
affect the concentration distribution, potentially leading to steady
anomalies. Therefore, the wavy structure of Jupiter's ammonia distribution
(Fig.~\ref{fig:intro}f-h) can be explained by the presence
of meridional circulation cells. On Jupiter, as condensation of ammonia
is expected only at the upper levels of the atmosphere ($0.5-1$~bar),
the ammonia concentration at those levels should be lower than at
depth \citep{fletcher2020}. In addition, precipitation, small-scale
turbulence, thermochemical and chemical reactions, and diffusion are
also expected to determine the vertical ammonia distribution ($M_{{\rm a}}$)
\citep{guillot2020a}. The $M_{{\rm a}}$ profile estimated from the
MWR \citep{Li2017} reveals a local minimum at $\sim6$~bar (Fig.~\ref{fig:theory}a).
This profile is used in this study as the background state, to explain
the ammonia anomalies.

Here, we focus on two regions with distinctly different deep dynamics:
the equatorial region (planetocentric latitudes $20^{\circ}{\rm S}$
to $20^{\circ}{\rm N}$), {where superrotation is assumed to be fueled
by eddy momentum fluxes perpendicular to the spin axis \citep[e.g.,][]{Busse2002}}, and midlatitudes
($60^{\circ}{\rm S}$ to $20^{\circ}{\rm S}$ and $20^{\circ}{\rm N}$
to $60^{\circ}{\rm N}$), where alternating jets are postulated to
be driven by horizontal eddies associated with mass-transporting meridional
cells \citep[e.g.,][]{Salyk2006,Schneider2009,young2019}.

We begin with the midlatitudes, where the meridional cells are mechanically
driven (see below) by turbulence, similar to Earth's Ferrel cells,
which form as a consequence of atmospheric waves breaking in midlatitudes
\citep{Vallis2006}. Unlike the largely baroclinic midlatitudes of
Earth, which result in mostly non-mass-transporting Ferrel cells \citep{juckes2001,Vallis2006},
the predominantly barotropic flows on Jupiter (at the depth range
associated with the MWR measurements) \citep{Kaspi2018,kaspi2020,galanti2020}
may allow mass-transporting meridional cells (see SI). {Consistently, deep convection models of Jupiter also show barotropic flows \citep[e.g.,][]{Busse1976,Aurnou2001}.} The upper branch
of Earth's Ferrel cells consists of a balance between the Coriolis
force and the eddy momentum flux convergence,

\begin{equation}
-f\bar{v}=-\frac{\partial\left(\overline{u^{'}v^{'}}\right)}{\partial y},\label{eq:ferrel_balance}
\end{equation}
where $f$ is the Coriolis parameter. This upper branch balance, which
is the leading order balance of the steady state zonal mean zonal
momentum equation, is expected to hold within the equivalent cells
on Jupiter (see SI). This balance can hold down to a depth of only
a few bars, as inferred from energy considerations \citep{Liu2010},
implying flows from belts to zones within the cloud layer of the Jovian
atmosphere. In the lower branch of the terrestrial Ferrel cells, the
balance is between the Coriolis force and a surface drag \citep{Vallis2006}.
{ Since the Jovian atmosphere lacks a bottom solid boundary, surface
drag cannot act to oppose the Coriolis force, although it has been
suggested that if the cells extend as deep as the jets \citep{kaspi2020},
the Lorentz force can act as a magnetic drag \citep{Liu2008,Liu2010,wicht2019}. Another possible mechanism that allows the jets to be barotropic in the upper atmosphere and decay in the interior is the presence of a stable layer, as was shown lately in several studies \citep{debras2019,christensen2020,wicht2020}}.

The direction of each Ferrel-like cell corresponds to the direction
of the respective midlatitude jet. Eastward (westward) jets are located
in cells of eastward momentum flux convergence (divergence) implying (Eq.~\ref{eq:ferrel_balance}) a counterclockwise (clockwise) circulation in the NH, and a
clockwise (counterclockwise) circulation in the SH (Fig.~\ref{fig:intro}b,d).
The upper branch of the Ferrel-like cells may coincide with the lower
branch of stacked upper cells with an opposite circulation \citep{Ingersoll2000,Showman2005}
(dashed transparent lines in Fig.~\ref{fig:theory}b), and therefore
may share the same balance (Eq.~\ref{eq:ferrel_balance}). Indications
for the upper cells come from temperature and shallow tracer distributions
\citep{Gierasch1986,fletcher2016,dePater2019}. Similar to the balance
describing the deeper branch of the lower cells, the upper branch
of the upper cells requires a drag force, which may result from breaking
of atmospheric waves \citep{Gierasch1986,ingersoll2021}.

\begin{figure}
\begin{centering}
\includegraphics[width=1\textwidth]{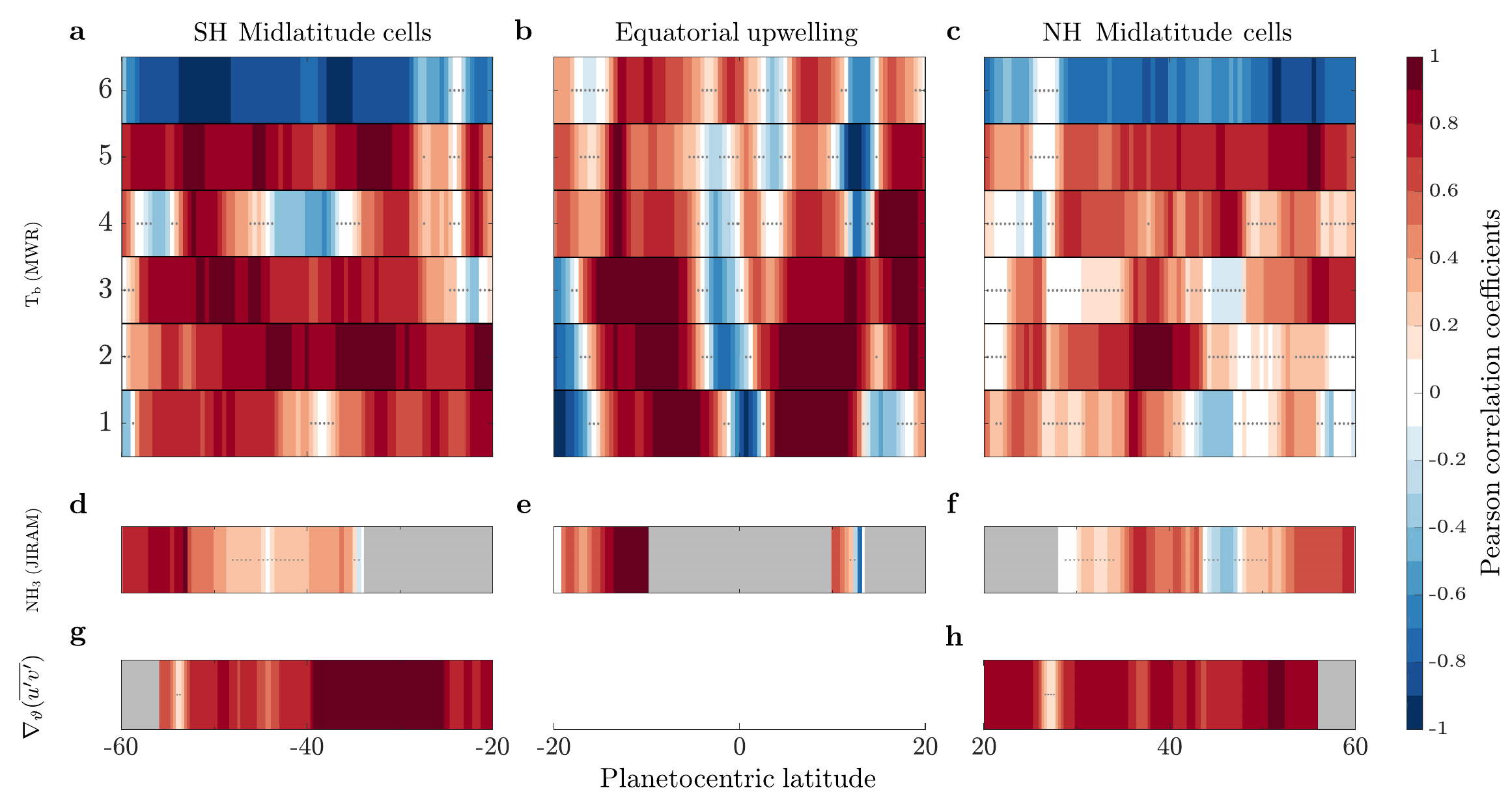}
\par\end{centering}
\caption{\label{fig:correlation}Pearson correlation coefficients as a function
of latitude. The correlations exemplify the relations presented in
Fig.~\ref{fig:theory}. (a) and (c), correlations
calculated between the zonal jets $\left(\bar{u}\right)$ and the
$T_{{\rm b}}$ meridional gradients $\left(\partial_{y}T_{{\rm b}}\right)$,
adjusted by the sign of the vertical gradient of $M_{{\rm a}}$, in
the six MWR channels, for the SH and NH, respectively. (b) Correlations
computed between the zonal jet velocity and $T_{{\rm b}}$ ($\bar{u}\propto-T_{{\rm b}}$).
(d) and (f), Correlations between the zonal velocity and the ammonia
abundance gradients ($\partial_{y}{\rm NH}_{3}$), measured by JIRAM,
in the SH and NH, respectively. (e) Correlations between
the zonal velocity and the ammonia abundance from JIRAM ($\bar{u}\propto{\rm NH}_{3}$).
(g) and (h) Correlations between the zonal velocity and the eddy momentum
flux convergence ($\bar{u}\propto-\partial_{y}\left(\overline{u^{\prime}v^{\prime}}\right)$),
in the SH and NH, respectively. Gray dots represent correlations that
are not statistically significant (confidence level $95\%$) and gray
regions show where measurements were not available. No data is available
for the eddy fluxes in the equatorial region, as evaluating them requires
measurements of the vertical winds (see SI), which are yet to be achieved.}
\end{figure}

The background ammonia profile is skewed by the vertical branches
of the cells (dashed orange lines in Fig.~\ref{fig:theory}b), maximizing
the ammonia meridional gradient where the jet velocity peaks (i.e.,
in the middle of the cell). This means that a correlation (along isobars) is expected between the zonal jets and the meridional gradient
of the ammonia concentration at midlatitudes \citep{duer2020,Fletcher2021}.
However, since the vertical gradient of $M_{{\rm a}}$ changes with
depth \citep{Li2017} (Fig.~\ref{fig:theory}a), the nature of the
correlation should change as well, as illustrated in Fig.~\ref{fig:theory}c,d.
These simple considerations motivate the examination of the correlation
between $\bar{u}$ and $\partial_{y}m_{a}$ $\left(-\partial_{y}T_{{\rm b}}\right)$
in midlatitudes (Fig.~\ref{fig:correlation}a,c, also see SI). Note
that $T_{{\rm b}}$ corresponds inversely to ammonia abundance at
a certain pressure level \citep{Li2017}. For a deep-wind estimate,
we use the measured cloud-level winds \citep{Tollefson2017} projected
inward in a direction parallel to the axis of rotation, without any
change in magnitude (in the upper $240$~bar), as implied by gravity measurement constraints
\citep{galanti2020,galanti2021}. The correlations are performed using
a $4^{\circ}$ latitudinal bin (see SI). By this, the suggested correlations
in Fig.~\ref{fig:theory} can be tested locally, rather than over
an entire hemisphere \citep{duer2020,Fletcher2021}.

At midlatitudes the overall positive correlations for MWR channels
1-5 indicate the existence of Ferrel-like cells at depths between
$1.5$ and $240$~bar (Fig.~\ref{fig:correlation}a,c). The positive
correlation with ammonia estimates by JIRAM (Fig.~\ref{fig:correlation}d,f)
further strengthens the prominence of the proposed cells. Channel
6 ($\sim0.7$~bar) exhibits negative correlations in midlatitudes
(Fig.~\ref{fig:correlation}a,c), implying that the deep cells do
not extend higher than $\sim1$~bar, and support the existence of
counter-rotating cells above that level \citep{Ingersoll2000,Showman2005,fletcher2020}.
To verify that the relation shown in Eq.~\ref{eq:ferrel_balance}
holds in the cells as illustrated in Fig.~\ref{fig:theory}b, regional
correlations between $\bar{u}$ and $-\partial_{y}\left(\overline{u^{'}v^{'}}\right)$
are shown for midlatitudes (Fig.~\ref{fig:correlation}g,h). This
positive correlation further strengthens the existence of the Ferrel-like
cells, where converging eddy momentum fluxes are the source of momentum. 
Overall, the correlation analysis reveals multiple deep Ferrel-like
cells, extending from $\sim1$~bar to at least $240$~bar.

The equatorial region of Jupiter, characterized by a wide eastward
jet, needs to be treated differently. Gravity analysis reveals that
Jupiter's interior (deeper than $\sim3000\,\,{\rm km}$) is rotating
as a rigid body \citep{Guillot2018}. Extending the zonal wind along
the direction of the spin axis thus separates the equatorial region
($17^{\circ}\,{\rm S}$ to $17^{\circ}\,{\rm N}$) from the truncated
cells at midlatitudes (see SI). The superrotating wind at low latitudes
requires a source of momentum \citep{imamura2020}. {Theories for such
sources include meridional \citep{potter2014,laraia2015} and vertical
\citep{Aurnou2001,Busse2002,Christensen2002,Heimpel2005,Kaspi2009,Dietrich2018}, propagation of waves. For the vertical case, several
studies have shown that an equatorial superrotation in giant planets
can be driven by eddy momentum fluxes perpendicular to the axis of
rotation \citep{Heimpel2005,Kaspi2009,Gastine2014}}. These fluxes
transfer momentum outwards and lead to a mean upwelling at the equatorial
region (see SI). Such an upwelling should lead to a maximum concentration
anomaly of any stably stratified matter (Fig.~\ref{fig:theory}e).
In the equatorial region, the minimum in $M_{{\rm a}}$ around $\sim6$
bar nearly vanishes (Fig.~S8), suggesting that the positive ammonia
anomalies at the equatorial region (Fig.~S2a) are due to the upwelling
from deep. To examine this, at the equator, the correlation is calculated
between the zonal velocity ($\bar{u}$) and the ammonia concentration
itself ($-T_{{\rm b}}$ for the MWR or ${\rm NH_{3}}$ for the JIRAM
measurements). Using a regional correlation analysis (see SI), it
is apparent that the correlations are largely positive at all depths
(Fig.~\ref{fig:correlation}b,e), implying that an equatorial upwelling
is dominant from the cloud deck and down to at least $240$~bar.
Very close to the equator, the correlation is negative due to the
local minimum in the zonal velocity (Fig.~\ref{fig:intro}b).

\section{Model reconstruction of Jupiter's ammonia distribution\label{sec:model}}

To further validate that the positive correlations shown in Fig.\ \ref{fig:correlation}a,c
are indeed due to the existence of meridional circulation cells, we
reconstruct the measured variations using a simplified advection-relaxation
model. Beginning with a steady-state zonal-mean conservation of species
equation for ammonia, assuming that diffusion terms are small, the
leading-order balance is

\begin{equation}
\bar{w}\left(\vartheta,r\right)\frac{\partial m_{a}\left(\vartheta,r\right)}{\partial r}+\bar{v}\left(\vartheta,r\right)\frac{\partial m_{{a}}\left(\vartheta,r\right)}{r\partial\vartheta}=-G\left(r\right)\left(m_{a}\left(\vartheta,r\right)-M_{{\rm a}}\left(r\right)\right),\label{eq: conservation of species-2}
\end{equation}
where $\bar{w}$ is the zonally averaged radial velocity, and $m_{a}$,
the variable solved for by the model, is the (zonal-mean) molar fraction
of ammonia. $M_{{\rm a}}$ is the ammonia concentration averaged over
isobaric surfaces (Fig.~\ref{fig:theory}a), and $G$ is the inverse
of a Newtonian relaxation timescale. The two terms on the left-hand
side represent advection by the mean circulation, and the right-hand
side term is a source term parameterized as a simple Newtonian relaxation
of ammonia. This relaxation term is assumed to include all
the processes resulting in the observed $M_{{\rm a}}$ as it acts
against local anomalies toward this mean vertical structure. To qualitatively
illustrate how the Ferrel-like cells' footprint might appear in the
ammonia distribution map ($m_{a}$), we solve the advection-relaxation
balance shown in Eq.\ \ref{eq: conservation of species-2}, for the midlatitudes between $1.5$ and $240$~bar (see SI). As
the balance in Eq.\ \ref{eq: conservation of species-2} indicates,
it is assumed that the relaxation time scale $\left(G^{-1}\right)$
is such that the advection and relaxation terms balance each other.

\begin{figure}
\begin{centering}
\includegraphics[width=0.8\textwidth]{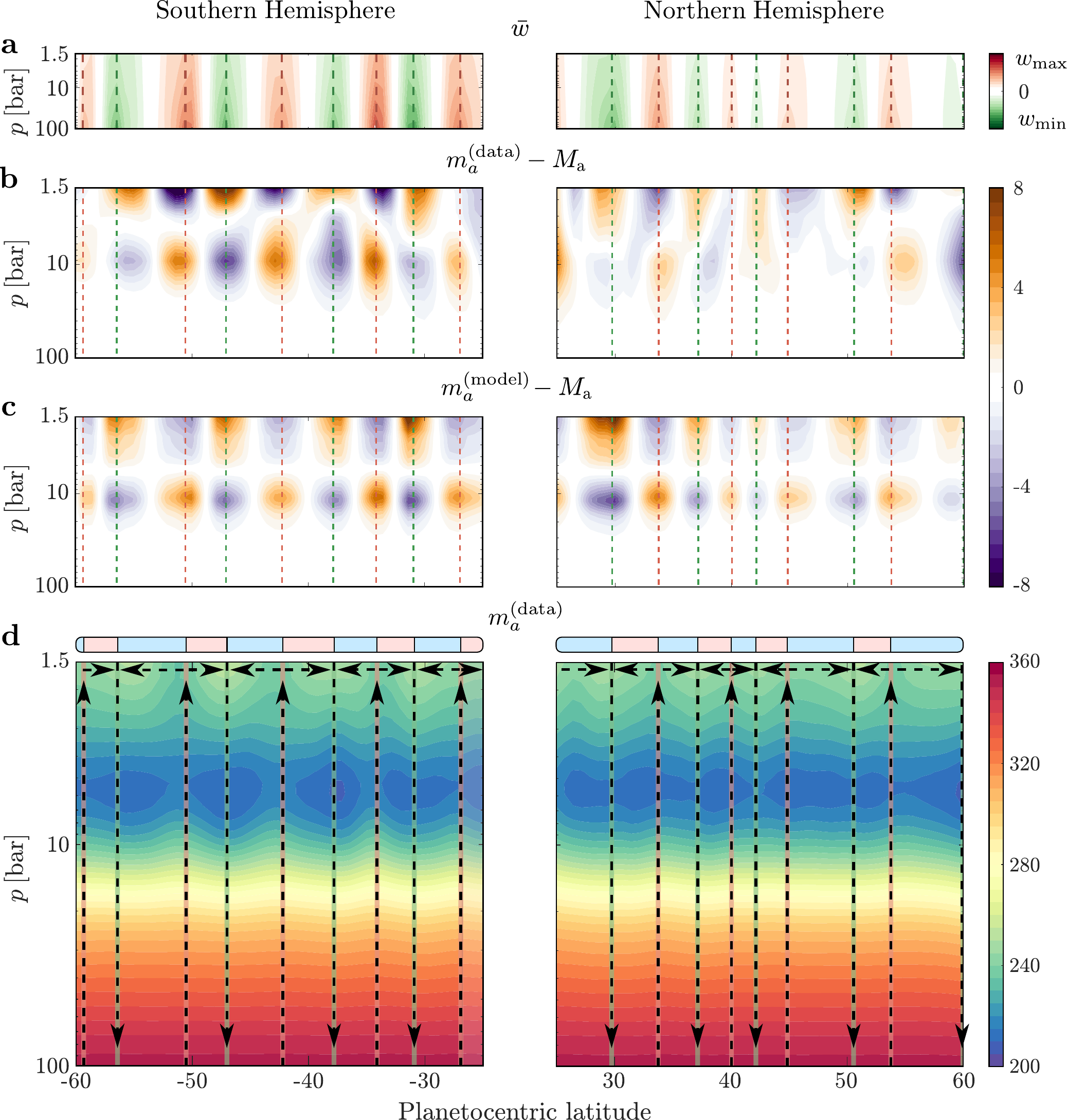}
\par\end{centering}
\caption{Jupiter's Ammonia distribution driven by an array of circulation cells.
(a) The normalized vertical zonal-mean wind ($\bar{w}$), as a function of latitude
and pressure, used in the model. Red and green contours are upward
and downward winds, respectively. (b) The ammonia anomalies reconstructed
from the data {[}ppm{]}. Here the vertical mean profile $M_{{\rm a}}$
is removed from the ammonia map $m_{a}$. (c) The ammonia anomalies
{[}ppm{]} produced by the advection-relaxation model. (d) The full
reconstructed ammonia map $m_{a}^{{\rm (data)}}$ {[}ppm{]}. Arrows
represent the direction of the cells' vertical and meridional winds.
(a-d) Red and green vertical lines are the locations of the upward
and downward branches of the cells, respectively. For reference, light
red (blue) bands indicate regions of positive (negative) vorticity
gradient as in Fig.~\ref{fig:intro}. The vertical axis is truncated
at $100$~bar as $M_{{\rm a}}$ becomes largely uniform beyond this
depth, thereby suppressing footprints of advection.\label{fig:model_results}}
\end{figure}

The zonally averaged velocity components $\left(\bar{v},\bar{w}\right)$
of the circulation cells, necessary for setting the advection terms
of Eq.~\ref{eq: conservation of species-2}, can be projected from
the available wind data according to the outline illustrated in Fig.~\ref{fig:theory}b.
Specifically, we relate between the circulation cells and the wind
data corresponding to the following assumptions (see also SI). The
borders between the cells are set at local extrema of the observed
cloud-level vorticity, the directions of the circulation cells are
set according to the directions of the jets in the middle of each
cell, and the strength of the circulation in each cell is set by the
measured eddy momentum flux convergence along the cell (Fig.~S5).
As the three terms in Eq.~\ref{eq: conservation of species-2} should
be proportional, but cannot be uniquely determined, the values of
$\bar{w}$, $\bar{v}$ and $G$ are normalized (Fig.\ \ref{fig:model_results}a).
This normalization means that while the model cannot produce absolute
values of winds due to unmeasured quantities, it can predict qualitatively
how these velocities would be structured spatially and what should
be their relative magnitudes, which are sufficient for assessing the
existence of the cells. Using scaling arguments, the value of Jupiter's static stability has recently been estimated to be in the order of $10^{-2} \; \rm{s^{-1}}$  \citep{Lee2021}, which can provide a further step towards estimating the magnitude of the velocities in the cells.

The described wind scheme results in upwellings
(downwellings) on the poleward (equatorward) sides of eastward jets
(Fig.\ \ref{fig:model_results}a). The cells are reversed for westward
jets. Finally, as a benchmark for the model results, derivation of
the ammonia abundance ($m_{a}^{({\rm data})}$) from the measured
$T_{{\rm b}}$ between the latitudes $60^{\circ}$S and $60^{\circ}$N
is implemented (see SI, Fig.~S2). As the depth of the cells, the
width of their branches and the parameter $G$ are unknown, an optimization
procedure is performed for determining these parameters to best match
the data (see SI). To ensure that this procedure does not influence
the qualitative nature of the results, Eq.~\ref{eq: conservation of species-2}
is also solved with a predefined physically-oriented set of parameters
(Figs.~S6 and S7).

Using the above assumptions, we solve Eq.~\ref{eq: conservation of species-2}
to predict the ammonia map ($m_{a}^{({\rm model})}$), and compare
it to $m_{a}^{({\rm data})}$ (Fig.~\ref{fig:model_results}). We
stress that the latitudinal variations appearing in the results (Fig.~\ref{fig:model_results}c),
stem only from the cloud-level wind observations without any assumption
on the meridional ammonia variation. For a clear comparison between
the $m_{a}^{({\rm data})}$ and $m_{a}^{({\rm model})}$, $M_{{\rm a}}$
is subtracted from both, such that only anomalies are visible (Fig.~\ref{fig:model_results}b,c).
Around 10~bar (Fig.\ \ref{fig:model_results}b), where $M_{{\rm a}}$
greatly increases with depth, enriched (depleted) ammonia anomalies
appear where upwellings (downwellings) are expected (Fig.~\ref{fig:model_results}a).
These features flip sign around the $6$-bar level, where $M_{{\rm a}}$
decreases with depth. These elements are captured well by the advection-relaxation
model (Fig.\ \ref{fig:model_results}c). In the SH, all $18$ anomalies
apparent in the observations have a counterpart of similar sign, shape
and position in the model results, suggesting the existence of 8 meridional
circulation cells. This agreement validates that advection by the
vertical branches of the cells is the main contributor in the creation
of the observed ammonia anomalies. {In the NH, similar results are
achieved, although the cells are slightly less coherent, perhaps due
to unexplained differences between the perijoves in the NH midlatitudes [Fig.~S3, 
\citet{oyafuso2020,Fletcher2021}], which might mask the cells' footprints
in the MWR data}. {Nevertheless, the lightning data
reinforces the existence of the NH cells, as lightning peaks are aligned
with the rising branch of the cells at the poleward side of the eastward
jets (Fig.\ \ref{fig:intro}e, Fig.~S1)}, which combined with the MWR data
(Fig.\ \ref{fig:model_results}b-d) provide indication for 8 northern
cells. Additional NH centered perijoves during the Juno extended mission
may provide data to better constrain the NH cells. For more intuition,
one can look at the full ammonia map (Fig.\ \ref{fig:model_results}d),
where iso-concentration lines are pulled up and down by the vertical
winds (as schematically illustrated in Fig.\ \ref{fig:theory}b),
emphasizing the locations of the 16 eddy-driven cells evident in the
MWR data.

\section{Discussion}

\begin{figure}
\begin{centering}
\includegraphics[width=0.8\textwidth]{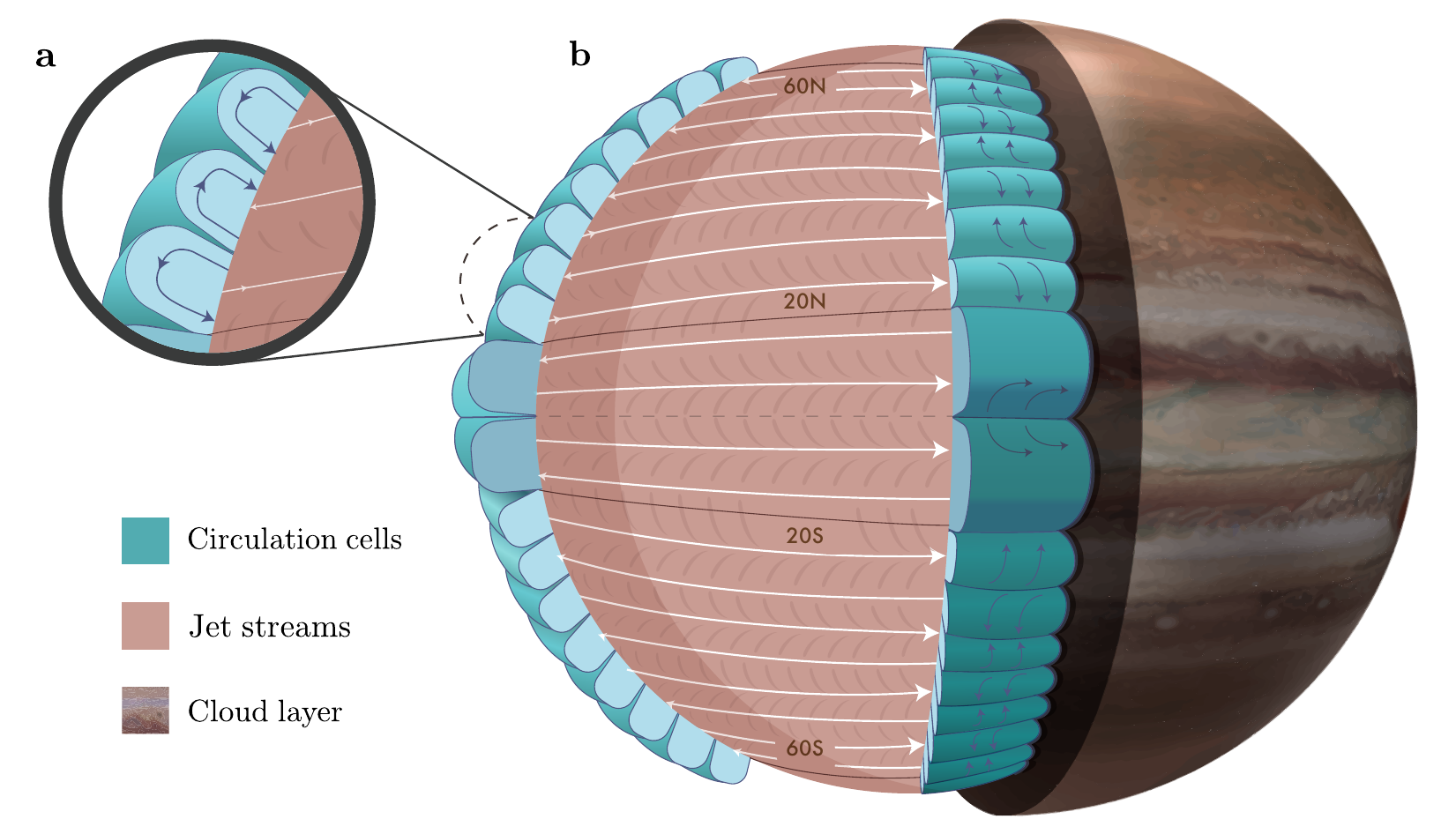}
\par\end{centering}
\caption{\label{fig:summary} (b) A figurative
cross section of Jupiter's meridional circulation and (a) a magnification
of the midlatitude circulation cells. The circulation cells (blue)
are axisymmetric in the zonal direction. The pink shell represents
a deep layer characteristic for all depths within the circulation
cells. The white arrows represent alternating jet streams and are
symmetric around the equator for the purpose of clarity. Each jet
between latitudes $20^{\circ}-60^{\circ}\,{\rm S/N}$ is accompanied
by a turbulence-driven circulation cell (blue arrows) in the meridional
plane as illustrated in panel b. The equatorial upwelling associated
with the superrotating jet is drawn at the equator, as illustrated
in panel e, as part of a larger possible equatorial cell (dark blue).}
\end{figure}

The identified array of alternating cells in midlatitudes, along with
the equatorial upwelling, are key features in the meridional overturning
circulation of the Jovian atmosphere (Fig.~\ref{fig:summary}).  {The cell's depth that can be inferred from the MWR measurements is limited to the sensing range ($\sim 240$~bar), and while the midlatitudinal cells are mechanically driven, as the Ferrel cells in Earth's troposphere, they are likely to extend deeper into the planet, as suggested by multiple theoretical studies \citep[e.g.,][]{Liu2010,christensen2020}. Similarly, deep meridional cells, which are mechanically driven, have been suggested to exist on the Sun \citep{miesch2011}.} 

This study provides an explanation for the observed meridional ammonia
anomalies, given the meridionally averaged vertical ammonia profile. The consistency of these results suggest that the $T_{{\rm b}}$ latitudinal variations are dominated by the opacity of a passive tracer, rather than the kinetic temperature. Note that evidence for the part of the deep cells extending from $1.5$
to $6$~bar depends on the flip of the background ammonia gradient (Fig.~\ref{fig:theory}a), and without
it these depths might be part of upper inverse cells \citep{Fletcher2021}. The shape of this vertical profile might be set by precipitation,
diffusion, and small-scale mixing, all of which might change with
latitude and depth \citep{guillot2020b}. Nonetheless, the remarkable
agreement between the model and the data, together with the robust
correlation analysis, provide strong evidence that the observed distribution
of ammonia is governed by the existence, number, position and relative
strength of the Ferrel-like circulation cells in Jupiter.

\subsection*{acknowledgments}
All the data used
in this study is publicly available, see \citet{Tollefson2017} for the winds data, \citet{Salyk2006} for the eddies data, \citet{Li2017} for the ammonia data, \citet{brown2018} for the lightning data and \citet{oyafuso2020} for the brightness temperature data.

\bibliographystyle{apalike}

\begin{thebibliography}{}

\bibitem[Aurnou and Olson, 2001]{Aurnou2001}
Aurnou, J.~M. and Olson, P.~L. (2001).
\newblock Strong zonal winds from thermal convectionin a rotating spherical
  shell.
\newblock {\em Geophys. Res. Lett.}, 28(13):2557--2559.

\bibitem[{Bolton} et~al., 2017]{Bolton2017}
{Bolton}, S.~J., {Adriani}, A., {Adumitroaie}, V., {Allison}, M., {Anderson},
  J., {Atreya}, S., {Bloxham}, J., {Brown}, S., {Connerney}, J.~E.~P.,
  {DeJong}, E., {Folkner}, W., {Gautier}, D., {Grassi}, D., {Gulkis}, S.,
  {Guillot}, T., {Hansen}, C., {Hubbard}, W.~B., {Iess}, L., {Ingersoll}, A.,
  {Janssen}, M., {Jorgensen}, J., {Kaspi}, Y., {Levin}, S.~M., {Li}, C.,
  {Lunine}, J., {Miguel}, Y., {Mura}, A., {Orton}, G., {Owen}, T., {Ravine},
  M., {Smith}, E., {Steffes}, P., {Stone}, E., {Stevenson}, D., {Thorne}, R.,
  {Waite}, J., {Durante}, D., {Ebert}, R.~W., {Greathouse}, T.~K., {Hue}, V.,
  {Parisi}, M., {Szalay}, J.~R., and {Wilson}, R. (2017).
\newblock {J}upiter's interior and deep atmosphere: The initial pole-to-pole
  passes with the {J}uno spacecraft.
\newblock {\em Science}, 356:821--825.

\bibitem[Brown et~al., 2018]{brown2018}
Brown, S., Janssen, M., Adumitroaie, V., Atreya, S., Bolton, S., Gulkis, S.,
  Ingersoll, A., Levin, S., Li, C., Li, L., Lunine, J., Misra, S., Orton, G.,
  Steffes, P., Tabataba-Vakili, F., Kolmasova, I., Imai, M., Santolik, O.,
  Kurth, W., Hospodarsky, G., Gurnett, D., and Connerney, J. (2018).
\newblock Prevalent lightning sferics at 600 megahertz near {J}upiter{'}s
  poles.
\newblock {\em Nature}, 558(7708):87--90.

\bibitem[Busse, 1976]{Busse1976}
Busse, F.~H. (1976).
\newblock A simple model of convection in the {J}ovian atmosphere.
\newblock {\em Icarus}, 29:255--260.

\bibitem[{Busse}, 2002]{Busse2002}
{Busse}, F.~H. (2002).
\newblock Convective flows in rapidly rotating spheres and their dynamo action.
\newblock {\em Phys. of Fluids.}, 14:1301--1314.

\bibitem[Christensen, 2002]{Christensen2002}
Christensen, U.~R. (2002).
\newblock Zonal flow driven by strongly supercritical convection in rotating
  spherical shells.
\newblock {\em J. Comp. Phys.}, 470:115--133.

\bibitem[Christensen et~al., 2020]{christensen2020}
Christensen, U.~R., Wicht, J., and Dietrich, W. (2020).
\newblock Mechanisms for limiting the depth of zonal winds in the gas giant
  planets.
\newblock {\em Astrophys. J.}, 890(1):61.

\bibitem[{de Pater} et~al., 2001]{DePater2001}
{de Pater}, I., {Dunn}, D., {Romani}, P., and {Zahnle}, K. (2001).
\newblock {Reconciling {G}alileo Probe Data and Ground-Based Radio Observations
  of Ammonia on {J}upiter}.
\newblock {\em Icarus}, 149(1):66--78.

\bibitem[de~{P}ater et~al., 2019]{dePater2019}
de~{P}ater, I., Sault, R.~J., Wong, M.~H., Fletcher, L.~N., De{B}oer, D., and
  Butler, B. (2019).
\newblock Jupiter{'}s ammonia distribution derived from {VLA} maps at 3-37
  {GH}z.
\newblock {\em Icarus}, 322:168--191.

\bibitem[Debras and Chabrier, 2019]{debras2019}
Debras, F. and Chabrier, G. (2019).
\newblock New models of {J}upiter in the context of {J}uno and {G}alileo.
\newblock {\em Astrophys. J.}, 872(1):100.

\bibitem[{Dietrich} and {Jones}, 2018]{Dietrich2018}
{Dietrich}, W. and {Jones}, C.~A. (2018).
\newblock {Anelastic spherical dynamos with radially variable electrical
  conductivity}.
\newblock {\em Icarus}, 305:15--32.

\bibitem[Duer et~al., 2020]{duer2020}
Duer, K., Galanti, E., and Kaspi, Y. (2020).
\newblock The range of {J}upiter{'}s flow structures that fit the {J}uno
  asymmetric gravity measurements.
\newblock {\em J. Geophys. Res. (Planets)}, 125(8).

\bibitem[Fletcher et~al., 2016]{fletcher2016}
Fletcher, L.~N., Greathouse, T.~K., Orton, G.~S., Sinclair, J.~A., Giles,
  R.~S., Irwin, P.~G., and Encrenaz, T. (2016).
\newblock Mid-infrared mapping of {J}upiter{'}s temperatures, aerosol opacity
  and chemical distributions with {IRTF/TEXES}.
\newblock {\em Icarus}, 278:128--161.

\bibitem[Fletcher et~al., 2020]{fletcher2020}
Fletcher, L.~N., Kaspi, Y., Guillot, T., and Showman, A.~P. (2020).
\newblock How well do we understand the belt/zone circulation of {G}iant
  {P}lanet atmospheres?
\newblock {\em Space Sci. Rev.}, 216(2):1--33.

\bibitem[Fletcher et~al., 2021]{Fletcher2021}
Fletcher, L.~N., Oyafuso, F.~A., Allison, M., Ingersoll, A., Li, L., Kaspi, Y.,
  Galanti, E., Wong, M.~H., Orton, G.~S., Duer, K., Zhang, Z., Li, C., Guillot,
  T., Levin, S.~M., and Bolton, S. (2021).
\newblock Jupiter's temperate belt/zone contrasts revealed at depth by {J}uno
  microwave observations.
\newblock {\em Earth and Space Science Open Archive}, page~35.

\bibitem[Galanti and Kaspi, 2021]{galanti2020}
Galanti, E. and Kaspi, Y. (2021).
\newblock Combined magnetic and gravity measurements probe the deep zonal flows
  of the gas giants.
\newblock {\em Mon. Not. Roy. Astro. Soc.}, 501(2):2352--2362.

\bibitem[Galanti et~al., 2021]{galanti2021}
Galanti, E., Kaspi, Y., Duer, K., Fletcher, L.~N., Ingersoll, A., Cheng, L.,
  Orton, G.~S., Guillot, T., M., L.~S., and J., B.~S. (2021).
\newblock Constraints on the latitudinal profile of {J}upiter{'}s deep jets.
\newblock {\em Geophys. Res. Lett.}, 48(9):e2021GL092912.

\bibitem[Garc{\i}a-Melendo and S{\'a}nchez-Lavega, 2001]{garcia2001}
Garc{\i}a-Melendo, E. and S{\'a}nchez-Lavega, A. (2001).
\newblock A study of the stability of jovian zonal winds from {HST} images:
  1995--2000.
\newblock {\em Icarus}, 152(2):316--330.

\bibitem[{Gastine} et~al., 2014]{Gastine2014}
{Gastine}, T., {Wicht}, J., {Duarte}, L.~D.~V., {Heimpel}, M., and {Becker}, A.
  (2014).
\newblock {Explaining {J}upiter's magnetic field and equatorial jet dynamics}.
\newblock {\em Geophys. Res. Lett.}, 41:5410--5419.

\bibitem[{Gierasch} et~al., 1986]{Gierasch1986}
{Gierasch}, P.~J., {Magalhaes}, J.~A., and {Conrath}, B.~J. (1986).
\newblock Zonal mean properties of {J}upiter's upper troposphere from {V}oyager
  infrared observations.
\newblock {\em Icarus}, 67:456--483.

\bibitem[Giles et~al., 2017]{giles2017}
Giles, R.~S., Fletcher, L.~N., Irwin, P.~G., Orton, G.~S., and Sinclair, J.~A.
  (2017).
\newblock Ammonia in {J}upiter{'}s troposphere from high-resolution 5 $\mu$m
  spectroscopy.
\newblock {\em Geophys. Res. Lett.}, 44(21):10--838.

\bibitem[Grassi et~al., 2010]{grassi2010}
Grassi, D., Adriani, A., Moriconi, M.~L., Ignatiev, N.~I., D'Aversa, E.,
  Colosimo, F., Negr{\~a}o, A., Brower, L., Dinelli, B.~M., Coradini, A., and
  Iccioni, G. (2010).
\newblock Jupiter{'}s hot spots: {Q}uantitative assessment of the retrieval
  capabilities of future {IR} spectro-imagers.
\newblock {\em Planatary and Space Science}, 58(10):1265--1278.

\bibitem[Grassi et~al., 2020]{grassi2020}
Grassi, D., Adriani, A., Mura, A., Atreya, S.~K., Fletcher, L.~N., Lunine,
  J.~I., Orton, G.~S., Bolton, S., Plainaki, C., Sindoni, G., Altieri, F.,
  Cicchetti, A., Dinelli, B.~M., Filacchione, G., Migliorini, A., Moriconi,
  M.~L., Noschese, R., Olivieri, A., Piccioni, G., Sordini, R., Stefani, S.,
  Tosi, F., and Turrini, D. (2020).
\newblock On the spatial distribution of minor species in {J}upiter{'}s
  troposphere as inferred from {J}uno {JIRAM} data.
\newblock {\em J. Geophys. Res. (Planets)}, 125(4).

\bibitem[Guillot et~al., 2020a]{guillot2020b}
Guillot, T., Li, C., Bolton, S.~J., Brown, S.~T., Ingersoll, A.~P., Janssen,
  M.~A., Levin, S.~M., Lunine, J.~I., Orton, G.~S., Steffes, P.~G., and
  Stevenson, D.~J. (2020a).
\newblock Storms and the depletion of ammonia in {J}upiter: Ii. explaining the
  {J}uno observations.
\newblock {\em J. Geophys. Res. (Planets)}, 125(8):e2020JE006404.

\bibitem[{Guillot} et~al., 2018]{Guillot2018}
{Guillot}, T., {Miguel}, Y., {Militzer}, B., {Hubbard}, W.~B., {Kaspi}, Y.,
  {Galanti}, E., {Cao}, H., {Helled}, R., {Wahl}, S.~M., {Iess}, L., {Folkner},
  W.~M., {Stevenson}, D.~J., {Lunine}, J.~I., {Reese}, D.~R., {Biekman}, A.,
  {Parisi}, M., {Durante}, D., {Connerney}, J.~E.~P., {Levin}, S.~M., and
  {Bolton}, S.~J. (2018).
\newblock A suppression of differential rotation in {J}upiter{'}s deep
  interior.
\newblock {\em Nature}, 555:227--230.

\bibitem[Guillot et~al., 2020b]{guillot2020a}
Guillot, T., Stevenson, D.~J., Atreya, S.~K., Bolton, S.~J., and Becker, H.~N.
  (2020b).
\newblock Storms and the depletion of ammonia in {J}upiter: I. microphysics of
  mushballs.
\newblock {\em J. Geophys. Res. (Planets)}, 125(8):e2020JE006403.

\bibitem[{Heimpel} et~al., 2005]{Heimpel2005}
{Heimpel}, M., {Aurnou}, J., and {Wicht}, J. (2005).
\newblock Simulation of equatorial and high-latitude jets on {J}upiter in a
  deep convection model.
\newblock {\em Nature}, 438:193--196.

\bibitem[{Iess} et~al., 2018]{Iess2018}
{Iess}, L., {Folkner}, W.~M., {Durante}, D., {Parisi}, M., {Kaspi}, Y.,
  {Galanti}, E., {Guillot}, T., {Hubbard}, W.~B., {Stevenson}, D.~J.,
  {Anderson}, J.~D., {Buccino}, D.~R., {Casajus}, L.~G., {Milani}, A., {Park},
  R., {Racioppa}, P., {Serra}, D., {Tortora}, P., {Zannoni}, M., {Cao}, H.,
  {Helled}, R., {Lunine}, J.~I., {Miguel}, Y., {Militzer}, B., {Wahl}, S.,
  {Connerney}, J.~E.~P., {Levin}, S.~M., and {Bolton}, S.~J. (2018).
\newblock Measurement of {J}upiter's asymmetric gravity field.
\newblock {\em Nature}, 555(7695):220--222.

\bibitem[Imamura et~al., 2020]{imamura2020}
Imamura, T., Mitchell, J., Lebonnois, S., Kaspi, Y., Showman, A.~P., and
  Korablev, O. (2020).
\newblock Superrotation in planetary atmospheres.
\newblock {\em Space Sci. Rev.}, 216(5):1--41.

\bibitem[Ingersoll et~al., 2017]{ingersoll2017b}
Ingersoll, A.~P., Adumitroaie, V., Allison, M.~D., Atreya, S., Bellotti, A.~A.,
  Bolton, S.~J., Brown, S.~T., Gulkis, S., Janssen, M.~A., Levin, S.~M., Cheng,
  L., Liming, L., Lunine, J.~I., Orton, G.~S., Oyafuso, F.~A., and Steffes,
  P.~G. (2017).
\newblock Implications of the ammonia distribution on {J}upiter from 1 to 100
  bars as measured by the {J}uno microwave radiometer.
\newblock {\em Geophys. Res. Lett.}, 44(15):7676--7685.

\bibitem[Ingersoll et~al., 2021]{ingersoll2021}
Ingersoll, A.~P., Atreya, S., Bolton, S.~J., Brueschaber, S., Fletcher, L.~N.,
  Galanti, E., Kaspi, Y., Levin, S.~M., Li, C., Li, L., Lunine, J.~I., Orton,
  G.~S., and Waite, H. (2021).
\newblock Jupiter{'}s overturning circulation: Breaking waves take the place of
  solid boundaries.
\newblock {\em Geophys. Res. Lett.}
\newblock in review.

\bibitem[{Ingersoll} et~al., 2000]{Ingersoll2000}
{Ingersoll}, A.~P., {Gierasch}, P.~J., {Banfield}, D., {Vasavada}, A.~R., and
  {Galileo Imaging Team} (2000).
\newblock Moist convection as an energy source for the large-scale motions in
  {J}upiter's atmosphere.
\newblock {\em Nature}, 403:630--632.

\bibitem[{Janssen} et~al., 2017]{janssen2017}
{Janssen}, M.~A., {Oswald}, J.~E., {Brown}, S.~T., {Gulkis}, S., {Levin},
  S.~M., {Bolton}, S.~J., {Allison}, M.~D., {Atreya}, S.~K., {Gautier}, D.,
  {Ingersoll}, A.~P., {Lunine}, J.~I., {Orton}, G.~S., {Owen}, T.~C.,
  {Steffes}, P.~G., {Adumitroaie}, V., {Bellotti}, A., {Jewell}, L.~A., {Li},
  C., {Li}, L., {Misra}, S., {Oyafuso}, F.~A., {Santos-Costa}, D.,
  {Sarkissian}, E., {Williamson}, R., {Arballo}, J.~K., {Kitiyakara}, A.,
  {Ulloa-Severino}, A., {Chen}, J.~C., {Maiwald}, F.~W., {Sahakian}, A.~S.,
  {Pingree}, P.~J., {Lee}, K.~A., {Mazer}, A.~S., {Redick}, R., {Hodges},
  R.~E., {Hughes}, R.~C., {Bedrosian}, G., {Dawson}, D.~E., {Hatch}, W.~A.,
  {Russell}, D.~S., {Chamberlain}, N.~F., {Zawadski}, M.~S., {Khayatian}, B.,
  {Franklin}, B.~R., {Conley}, H.~A., {Kempenaar}, J.~G., {Loo}, M.~S.,
  {Sunada}, E.~T., {Vorperion}, V., and {Wang}, C.~C. (2017).
\newblock {{MWR}: {M}icrowave {R}adiometer for the {J}uno {M}ission to
  {J}upiter}.
\newblock {\em Space Sci. Rev.}, 213(1-4):139--185.

\bibitem[Juckes, 2001]{juckes2001}
Juckes, M. (2001).
\newblock A generalization of the transformed {E}ulerian-mean meridional
  circulation.
\newblock {\em Q. J. R. Meteorol. Soc.}, 127(571):147--160.

\bibitem[{Kaspi} et~al., 2009]{Kaspi2009}
{Kaspi}, Y., {Flierl}, G.~R., and {Showman}, A.~P. (2009).
\newblock The deep wind structure of the giant planets: Results from an
  anelastic general circulation model.
\newblock {\em Icarus}, 202:525--542.

\bibitem[{Kaspi} et~al., 2018]{Kaspi2018}
{Kaspi}, Y., {Galanti}, E., {Hubbard}, W.~B., {Stevenson}, D.~J., {Bolton},
  S.~J., {Iess}, L., {Guillot}, T., {Bloxham}, J., {Connerney}, J.~E.~P.,
  {Cao}, H., {Durante}, D., {Folkner}, W.~M., {Helled}, R., {Ingersoll}, A.~P.,
  {Levin}, S.~M., {Lunine}, J.~I., {Miguel}, Y., {Militzer}, B., {Parisi}, M.,
  and {Wahl}, S.~M. (2018).
\newblock {J}upiter's atmospheric jet-streams extend thousands of kilometres
  deep.
\newblock {\em Nature}, 555:223--226.

\bibitem[Kaspi et~al., 2020]{kaspi2020}
Kaspi, Y., Galanti, E., Showman, A.~P., Stevenson, D.~J., Guillot, T., Iess,
  L., and Bolton, S.~J. (2020).
\newblock Comparison of the deep atmospheric dynamics of {J}upiter and {S}aturn
  in light of the {J}uno and {C}assini gravity measurements.
\newblock {\em Space Sci. Rev.}, 216(5):1--27.

\bibitem[Laraia and Schneider, 2015]{laraia2015}
Laraia, A.~L. and Schneider, T. (2015).
\newblock Superrotation in terrestrial atmospheres.
\newblock {\em J. Atmos. Sci.}, 72(11):4281--4296.

\bibitem[Lee and Kaspi, 2021]{Lee2021}
Lee, S. and Kaspi, Y. (2021).
\newblock Towards an understanding of the structure of {J}upiter{'}s atmosphere
  using the ammonia distribution and the {T}ransformed {E}ulerian {M}ean
  theory.
\newblock {\em J. Atmos. Sci.}, 78(7):2047--2056.

\bibitem[Lewis et~al., 2007]{lewis2007}
Lewis, S.~R., Read, P.~L., Conrath, B.~J., Pearl, J.~C., and Smith, M.~D.
  (2007).
\newblock Assimilation of thermal emission spectrometer atmospheric data during
  the {M}ars {G}lobal {S}urveyor aerobraking period.
\newblock {\em Icarus}, 192(2):327--347.

\bibitem[Li et~al., 2020]{li2020}
Li, C., Ingersoll, A., Bolton, S., Levin, S., Janssen, M., Atreya, S., Lunine,
  J., Steffes, P., Brown, S., Guillot, T., Allison, M., Arballo, J., Bellotti,
  A., Adumitroaie, V., Gulkis, S., Hodges, A., Li, L., Misra, S., Orton, G.,
  Oyafuso, F., Santos-Costa, D., Waite, H., and Zhang, Z. (2020).
\newblock The water abundance in {J}upiter{'}s equatorial zone.
\newblock {\em Nature Astronomy}, 4(6):609--616.

\bibitem[Li et~al., 2017]{Li2017}
Li, C., Ingersoll, A., Janssen, M., Levin, S., Bolton, S., Adumitroaie, V.,
  Allison, M., Arballo, J., Bellotti, A., Brown, S., Ewald, S., Jewell, L.,
  Misra, S., Orton, G., Oyafuso, F., Steffes, P., and Williamson, R. (2017).
\newblock The distribution of ammonia on {J}upiter from a preliminary inversion
  of {J}uno microwave radiometer data.
\newblock {\em Geophys. Res. Lett.}, 44(11):5317--5325.

\bibitem[Limaye, 2007]{limaye2007}
Limaye, S.~S. (2007).
\newblock Venus atmospheric circulation: {K}nown and unknown.
\newblock {\em J. Geophys. Res. (Planets)}, 112(E4).

\bibitem[Little et~al., 1999]{little1999}
Little, B., Anger, C.~D., Ingersoll, A.~P., Vasavada, A.~R., Senske, D.~A.,
  Breneman, H.~H., Borucki, W.~J., and Team, T. G.~S. (1999).
\newblock {G}alileo images of lightning on {J}upiter.
\newblock {\em Icarus}, 142(2):306--323.

\bibitem[{Liu} et~al., 2008]{Liu2008}
{Liu}, J., {Goldreich}, P.~M., and {Stevenson}, D.~J. (2008).
\newblock {Constraints on deep-seated zonal winds inside {J}upiter and
  {S}aturn}.
\newblock {\em Icarus}, 196:653--664.

\bibitem[{Liu} and {Schneider}, 2010]{Liu2010}
{Liu}, J. and {Schneider}, T. (2010).
\newblock Mechanisms of jet formation on the giant planets.
\newblock {\em J. Atmos. Sci.}, 67:3652--3672.

\bibitem[Miesch and Hindman, 2011]{miesch2011}
Miesch, M.~S. and Hindman, B.~W. (2011).
\newblock Gyroscopic pumping in the solar near-surface shear layer.
\newblock {\em Astrophys. J.}, 743(1):79.

\bibitem[Oyafuso et~al., 2020]{oyafuso2020}
Oyafuso, F., Levin, S., Orton, G., Brown, S., Adumitroaie, V., Janssen, M.,
  Wong, M.~H., Fletcher, L.~N., Steffes, P., Li, C., Gulkis, S., Atreya, S.,
  S., M., and S., B. (2020).
\newblock Angular dependence and spatial distribution of {J}upiter{'}s
  centimeter-wave thermal emission from {J}uno{'}s microwave radiometer.
\newblock {\em Earth Planet. Sci. Lett.}, 7(11):e2020EA001254.

\bibitem[{Porco} et~al., 2003]{Porco2003}
{Porco}, C.~C., {West}, R.~A., {McEwen}, A., {Del Genio}, A.~D., {Ingersoll},
  A.~P., {Thomas}, P., {Squyres}, S., {Dones}, L., {Murray}, C.~D., {Johnson},
  T.~V., {Burns}, J.~A., {Brahic}, A., {Neukum}, G., {Veverka}, J., {Barbara},
  J.~M., {Denk}, T., {Evans}, M., {Ferrier}, J.~J., {Geissler}, P.,
  {Helfenstein}, P., {Roatsch}, T., {Throop}, H., {Tiscareno}, M., and
  {Vasavada}, A.~R. (2003).
\newblock {C}assini imaging of {J}upiter's atmosphere, satellites and rings.
\newblock {\em Science}, 299:1541--1547.

\bibitem[Potter et~al., 2014]{potter2014}
Potter, S.~F., Vallis, G.~K., and Mitchell, J.~L. (2014).
\newblock Spontaneous superrotation and the role of {K}elvin waves in an
  idealized dry {GCM}.
\newblock {\em J. Atmos. Sci.}, 71(2):596--614.

\bibitem[Read et~al., 2018]{read2018}
Read, P.~L., Lewis, S.~R., and Vallis, G.~K. (2018).
\newblock Atmospheric dynamics of terrestrial planets.
\newblock {\em Handbook of Exoplanets}, 144:2537--2557.

\bibitem[{Salyk} et~al., 2006]{Salyk2006}
{Salyk}, C., {Ingersoll}, A.~P., {Lorre}, J., {Vasavada}, A., and {Del Genio},
  A.~D. (2006).
\newblock Interaction between eddies and mean flow in {J}upiter's atmosphere:
  Analysis of {C}assini imaging data.
\newblock {\em Icarus}, 185:430--442.

\bibitem[{Schneider}, 2006]{Schneider2006a}
{Schneider}, T. (2006).
\newblock The general circulation of the atmosphere.
\newblock {\em Ann. Rev. Earth Plan. Sci.}, 34:655--688.

\bibitem[{Schneider} and {Liu}, 2009]{Schneider2009}
{Schneider}, T. and {Liu}, J. (2009).
\newblock Formation of jets and equatorial superrotation on {J}upiter.
\newblock {\em J. Atmos. Sci.}, 66:579--601.

\bibitem[{Showman} and {de Pater}, 2005]{Showman2005}
{Showman}, A.~P. and {de Pater}, I. (2005).
\newblock Dynamical implications of {J}upiter's tropospheric ammonia abundance.
\newblock {\em Icarus}, 174:192--204.

\bibitem[Taylor et~al., 2004]{taylor2004}
Taylor, F.~W., Atreya, S.~K., Encrenaz, T.~H., Hunten, D.~M., Irwin, P.~G., and
  Owen, T.~C. (2004).
\newblock {\em Jupiter: the planet, satellites and magnetosphere}, chapter The
  composition of the atmosphere of {J}upiter, pages 59--78.
\newblock Cambridge University Press.

\bibitem[Tollefson et~al., 2017]{Tollefson2017}
Tollefson, J., {Wong}, M.~H., {de Pater}, I., {Simon}, A.~A., {Orton}, G.~S.,
  {Rogers}, J.~H., {Atreya}, S.~K., C., R.~G., {Januszewski}, W.,
  {Morales-Juber{\'\i}as}, R., and S., M.~P. (2017).
\newblock Changes in {J}upiter's zonal wind profile preceding and during the
  {J}uno mission.
\newblock {\em Icarus}, 296:163--178.

\bibitem[{Vallis}, 2017]{Vallis2006}
{Vallis}, G.~K. (2017).
\newblock {\em Atmospheric and {O}ceanic {F}luid {D}ynamics}.
\newblock pp.~770.~Cambridge University Press., second edition.

\bibitem[{Vasavada} and {Showman}, 2005]{Vasavada2005}
{Vasavada}, A.~R. and {Showman}, A.~P. (2005).
\newblock {J}ovian atmospheric dynamics: {A}n update after {G}alileo and
  {C}assini.
\newblock {\em Reports of Progress in Physics}, 68:1935--1996.

\bibitem[Wicht and Gastine, 2020]{wicht2020}
Wicht, J. and Gastine, T. (2020).
\newblock Numerical simulations help revealing the dynamics underneath the
  clouds of {J}upiter.
\newblock {\em Nature Communications}, 11(1):1--4.

\bibitem[Wicht et~al., 2019]{wicht2019}
Wicht, J., Gastine, T., Duarte, L.~D., and Dietrich, W. (2019).
\newblock Dynamo action of the zonal winds in {J}upiter.
\newblock {\em Astron. and Astrophys.}, 629:A125.

\bibitem[Young and Read, 2017]{young2017}
Young, R.~M. and Read, P.~L. (2017).
\newblock Forward and inverse kinetic energy cascades in {J}upiter's turbulent
  weather layer.
\newblock {\em Nature Physics}, 13(11):1135--1140.

\bibitem[Young et~al., 2019]{young2019}
Young, R.~M., Read, P.~L., and Wang, Y. (2019).
\newblock Simulating {J}upiter{'}s weather layer. {P}art {I}: {J}et spin-up in
  a dry atmosphere.
\newblock {\em Icarus}, 326:225--252.

\end{thebibliography}

\newpage

\renewcommand{\thesection}{S\arabic{section}} 
\renewcommand{\thetable}{S\arabic{table}} 
\renewcommand{\thefigure}{S\arabic{figure}}
\renewcommand{\theequation}{S\arabic{equation}}
\setcounter{figure}{0}
\setcounter{section}{0}
\setcounter{equation}{0}

\part*{Supporting Information}

\section{Eddy momentum driven Ferrel-like cell}

\subsection{Standard formulation}

Using approximations similar to the commonly used formulation which
describe the terrestrial Ferrel-cell dynamics \citep{Vallis2006},
the leading order zonal mean zonal momentum equation may be written
as

\begin{equation}
\frac{\partial\bar{u}}{\partial t}+\frac{\partial}{\partial y}\left(\overline{u^{'}v^{'}}\right)-f\bar{v}=-F_{{\rm sink}},\label{eq:Ferrel-leading-order}
\end{equation}
where $F_{{\rm sink}}$ is a sink term. On Earth, the sink term represents
a surface drag in the Ekman layer and in the Jovian atmosphere, if
the cells are as deep as the jets \citep{kaspi2020}, it might be
Ohmic dissipation \citep{Liu2008,Liu2010,kaspi2020} (i.e., $F_{{\rm sink}}=\frac{1}{4\pi\bar{\rho}}\overline{\left(\nabla\times\mathbf{B}\right)\times\mathbf{B}},$
where ${\rm {\bf B}}$ is the 3D magnetic field and $\rho$ is density).
Eq.~1\footnote{Note that equation cross-references without a S refer to equations
in the main text.}, adequate for the upper branch of the cells, results from applying
time-averaging away from the sink layer. This balance leads to meridional
velocities in the directions illustrated in Fig.~2b\footnote{Note that figure cross-references without a S refer to figures in
the main text.}. Alternatively, applying vertical integration cancels the Coriolis
terms in the vertical boundaries of the cell and the relation between
the zonal jets, their eddy source term and the sink term becomes

\begin{equation}
\frac{\partial\bar{U}}{\partial t}=-\frac{\partial}{\partial y}\left(\overline{U^{'}V^{'}}\right)-\hat{F}_{{\rm sink}}.
\end{equation}
where $U$ and $V$ are the vertically integrated velocities, and
$\hat{F}_{{\rm sink}}$ is the vertically integrated sink term. Therefore,
converging (diverging) eddy momentum fluxes transfer their momentum
to eastward (westward) jets, as can be seen in the Jovian atmosphere
(Fig.~1b,d). In the descriptions throughout this study, Cartesian
approximations are used for the sake of clarity, but the actual calculations
were performed using the spherical, more accurate, formulations.

\subsection{TEM formulation for Jupiter}

The Transformed Eulerian Mean (TEM) equations, commonly invoked to
quantify Lagrangian mass-transport in Earth's Ferrel cells, describe
a circulation driven by the diabatic heating term \citep{Vallis2006}.
The TEM formulation can be derived from the momentum and thermodynamic
equations, under Boussinesq and quasi-geostrophic approximations (\citet{Vallis2006},
ch.~10.3), resulting in

\begin{equation}
\frac{\partial\bar{u}}{\partial t}-f\bar{v}^{*}=\nabla\cdot{ \mathcal{{F}}},\label{eq:TEM_momentum}
\end{equation}
and

\begin{equation}
\frac{\partial\bar{b}}{\partial t}-\bar{w}^{*}N^{2}=\bar{S},
\end{equation}
where $\bar{v}^{*}=\bar{v}-\frac{\partial}{\partial z}\left(\frac{1}{N^{2}}\overline{v'b'}\right)$
and $\bar{w}^{*}=\bar{w}+\frac{\partial}{\partial y}\left(\frac{1}{N^{2}}\overline{v'b'}\right)$
are defined as the ``residual'' mean meridional and vertical velocities,
respectively, which approximate mass-transport by both Eulerian mean
velocities and eddy fluxes, $N^{2}=\nicefrac{\partial\bar{b_{0}}}{\partial z}$
is the Brunt-Vaisala frequency, $\bar{S}$ is the diabatic heating
term, and ${\bf \mathcal{{F}}}=-\left(u'v'\right){{\rm \hat{j}}}+\left(\frac{f}{N^{2}}\overline{v'b'}\right){\rm {\hat{k}}}\,\,$
is the Elissan-Plam (EP) flux. \textbf{$b_{0}$ }and $b$ represent
the mean (zonally- and meridionally-averaged) and the deviation from
the mean of the buoyancy force and ${{\rm \hat{j}}}$ (${{\rm \hat{k}}}$)
is a unit vectors in the meridional (vertical) direction. On Earth,
the diabatic heating term is important \citep{Vallis2006}, and therefore
the residual meridional velocities accurately represent the total
meridional transport of mass in Earth's midlatitudes.

The midlatitude atmosphere on Earth is characterized by baroclinicity,
and as a result, the second term of the EP flux is substantial and
plays a key role in the resulting circulation. On Jupiter, the eddy
fluxes beneath the cloud level and the diabatic heating are yet to
be measured. However, gravity-measurement analysis implies that Jupiter's
jets are nearly barotropic in the depth range relevant to this study
\citep{galanti2020}, meaning that the EP flux is dominated by the
first term, and $\bar{w}^{*}$ is comparable to $\bar{w}$ \citep{Lee2021}.
Therefore, under the barotropic limit, the equations describing the
Eulerian velocities in a Ferrel-like cell might also represent the
total mass transport in the Jovian cells.

\subsection{Correlation analysis}

The picture illustrated in Fig.~2, relating the distribution of ammonia
and the zonal winds according to the Ferrel-like cells hypothesis,
is tested quantitatively in a correlation analysis exhibited in Fig.~3.
The expected relations between the zonal jets and the ammonia meridional
gradients in the NH are

\begin{equation}
\bar{u}\propto\begin{cases}
\begin{array}{c}
-\partial_{y}m_{a}\\
\partial_{y}m_{a}
\end{array} & \begin{array}{c}
1.5\leq p<6\,{\rm bar}\quad\left({\rm channels}\,\,4-5\right),\\
p\geq6\,{\rm bar}\quad\left({\rm channels\,}\,1-3\right),\,\,p<1.5\,{\rm bar}\quad\left({\rm channel\,}\,6\right).
\end{array}\end{cases}\label{eq:mid_lat}
\end{equation}

Here, $\partial_{y}m_{a}$ is the latitudinal gradient of the ammonia
concentration ($m_{a}$) and $p$ is pressure. The channels refer
to the six frequencies of Juno's MWR. In the SH, as the circulation
is reversed, Eq.~\ref{eq:mid_lat} flips signs. Eq.~\ref{eq:mid_lat}
captures also the case of cells with westward jets, as both $\bar{u}$
and $\partial_{y}m_{a}$ change sign. A Pearson correlation coefficient
(${{S}}(\vartheta,{\rm ch})$) is calculated for each latitude
and MWR channel, and its value is represented by a color between blue,
representing a negative correlation, white, representing no correlation
and red, representing a positive correlation. The $T_{{\rm b}}$ data
is measured in a resolution of $\sim0.6^{\circ}$ latitude. The data
is interpolated such that the grid size is $0.1^{\circ}$ latitude,
and the correlation for each point $\vartheta_{i}$ is calculated
along a span \{$\vartheta_{i}-2^{\circ}$,$\vartheta_{i}+2^{\circ}$\}.
This choice of a $4^{\circ}$ latitudinal bin allows having enough
data points for the statistical value of the correlation (more than
6 data pairs), and ensures the local nature of the results. The correlations
on the MWR data are calculated between the following trends. In Fig.~3a,
channels 1-3 and 6 and in Fig.~3c channels 4-5 the color represents
the correlation $\bar{u}\propto\partial_{y}T_{{\rm b}}$. In Fig.~3a,
channels 4-5 and in Fig.~3c channels 1-3 and 6 the color represents
the correlation $\bar{u}\propto-\partial_{y}T_{{\rm b}}$. Note that
anomalies of brightness temperature and ammonia abundance are inversely
proportional \citep{Li2017}. In Fig.~3b the correlations at all
channels are calculated according to $\bar{u}\propto-T_{{\rm b}}$.
Here, $T_{{\rm b}}$ is the Nadir component of the brightness temperate
{[}$^{\circ}{\rm K}${]} (other emission angles were not included
in the analysis), averaged over nine Juno orbits (PJs $1,3,4,5,6,7,8,9$
and $12$) \citep{oyafuso2020}. For further discussion regarding
limb-darkening $T_{{\rm b}}$ we point the readers to \citet{Fletcher2021}.
$\bar{u}$ is Jupiter's zonally-averaged zonal wind {[}${\rm m\,\,s^{-1}}${]}
measured by the Hubble space telescope during Juno's third perijove
\citep{Tollefson2017}, projected barotropically along the axis of
rotation \citep{galanti2020,galanti2021}. The ammonia distribution
by JIRAM is estimated at a depth of $\sim6$ bar, which is the depth
where a local minimum appears in $M_{{\rm a}}$ (Fig.~2a). Arbitrarily,
the correlation is performed according to $p>6$~bar in Eq.~2 (Fig.~2d)
and the overall positive result points that indeed the depth level
of JIRAM measurements should be deeper than the local minimum of $M_{{\rm a}}$.
The ammonia estimates from JIRAM are measured in a resolution of $1^{\circ}$
latitude. Similar to the $T_{{\rm b}}$ correlation analysis, the
data is interpolated, and the correlation is performed on a $4^{\circ}$
latitude bin for consistency. In Fig.~3d (f) the color represents
the correlation $\bar{u}\propto-\partial_{y}m_{a}$ ($\bar{u}\propto\partial_{y}m_{a}$)
and in Fig.~3e the color represents the correlation $\bar{u}\propto m_{a}$.
Finally, the eddy momentum flux convergence is measured in a resolution
of $1^{\circ}$ latitude, and the correlation is performed using the
same latitudinal bin of $4^{\circ}$. In Fig.~3g (h) the color represents
the correlation $\bar{u}\propto-\partial_{y}\left(\overline{u^{\prime}v^{\prime}}\right)$
($\bar{u}\propto\partial_{y}\left(\overline{u^{\prime}v^{\prime}}\right)$). 

\begin{figure}
\begin{centering}
\includegraphics[width=0.8\textwidth]{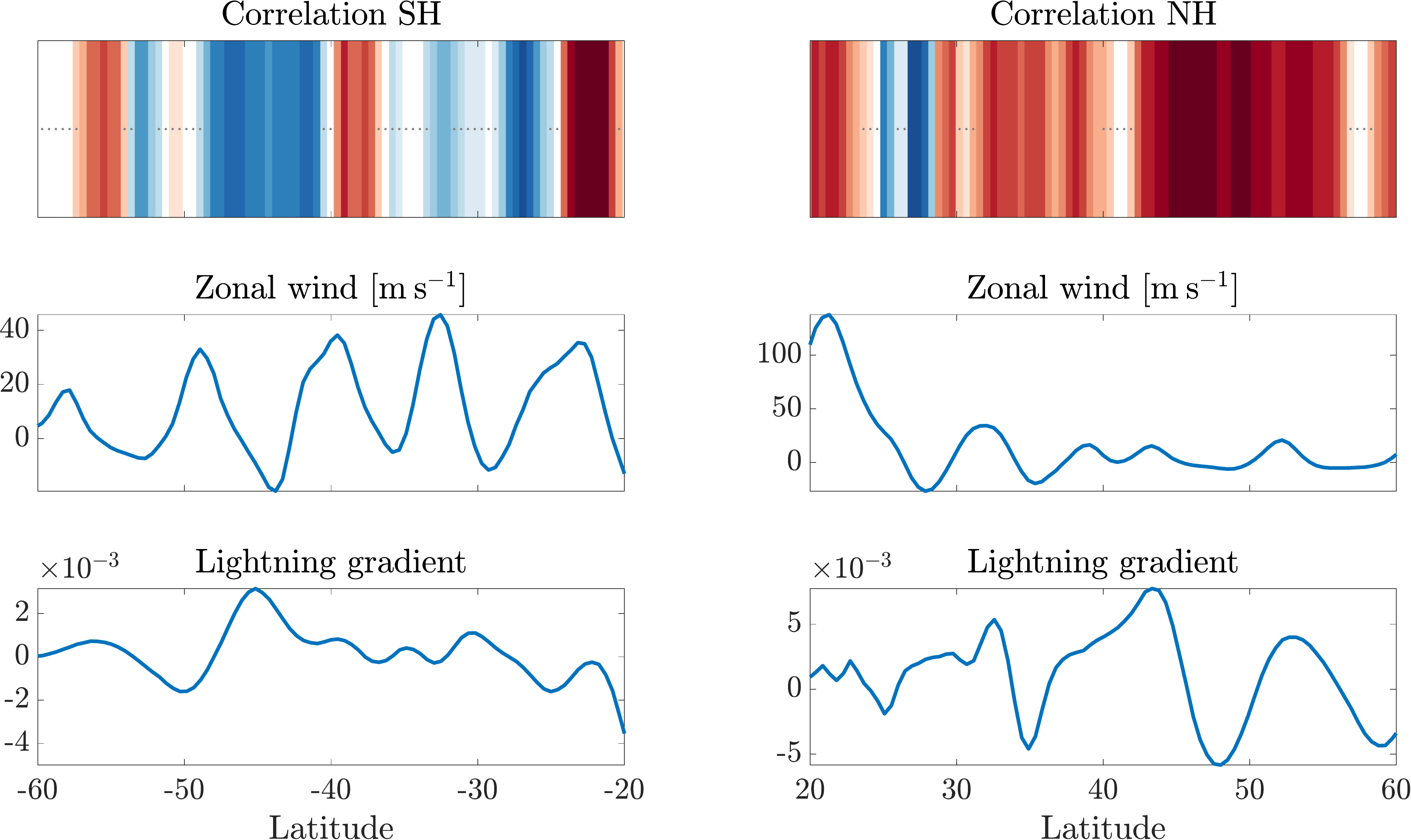}
\par\end{centering}
\caption{\label{fig:Lightning corr} Correlation coefficients (upper panels)
between the zonal wind (middle panels) and the lightning meridional
gradient (lower panels) in the midlatitudes.}
\end{figure}

We also examine the correlation between the zonal wind and the lightning
gradient. We find a good match in the northern hemisphere and a weak
negative correlation in the southern hemisphere, where Juno is much
less sensitive to lightnings (Fig.~\ref{fig:Lightning corr}). Note
that the correlation between lightnings and the Ferrel cells is less
indicative, as we should only examine the correlation in the rising
branch of the cells. Therefore, the correlation values away from the
rising branch should be regarded with caution.

\subsection{Advection-relaxation model}

\subsubsection{$T_{{\rm b}}$ as an indicator for ammonia}

\begin{figure}
\begin{centering}
\includegraphics[width=0.8\textwidth]{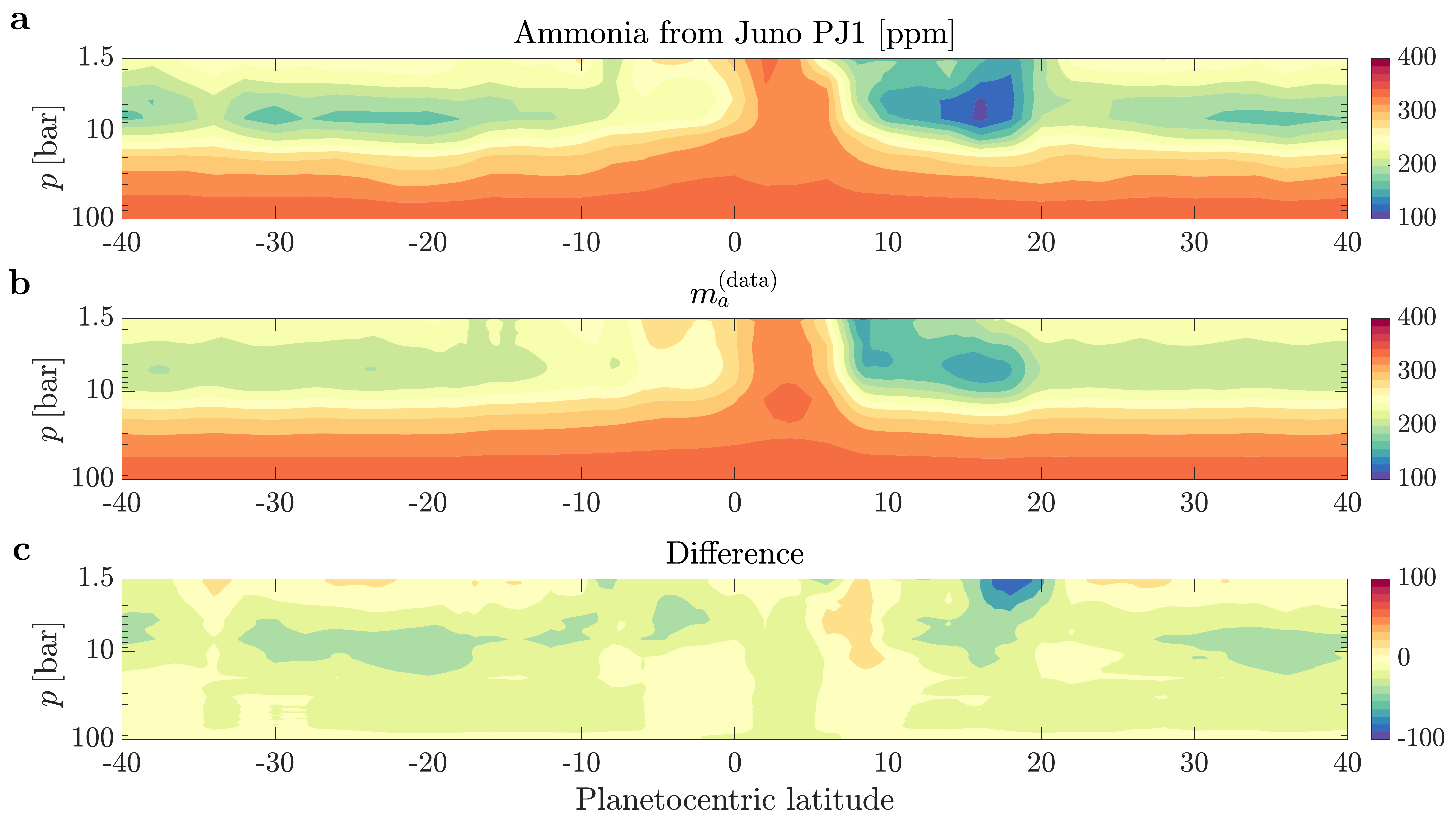}
\par\end{centering}
\caption{\label{fig:pseudo_NH3}(a)\textbf{ }Ammonia estimate \citep{Li2017}
from Juno MWR PJ1 data. (b) $m_{a}$ calculated from $T_{{\rm b}}$
(Eq.~\ref{eq:measured_ma}). This field is used as a benchmark ($m_{a}^{({\rm data})}$)
for the model results. (c) the difference between panel a and panel
b.}
\end{figure}
\begin{figure}
\begin{centering}
\includegraphics[width=0.7\textwidth]{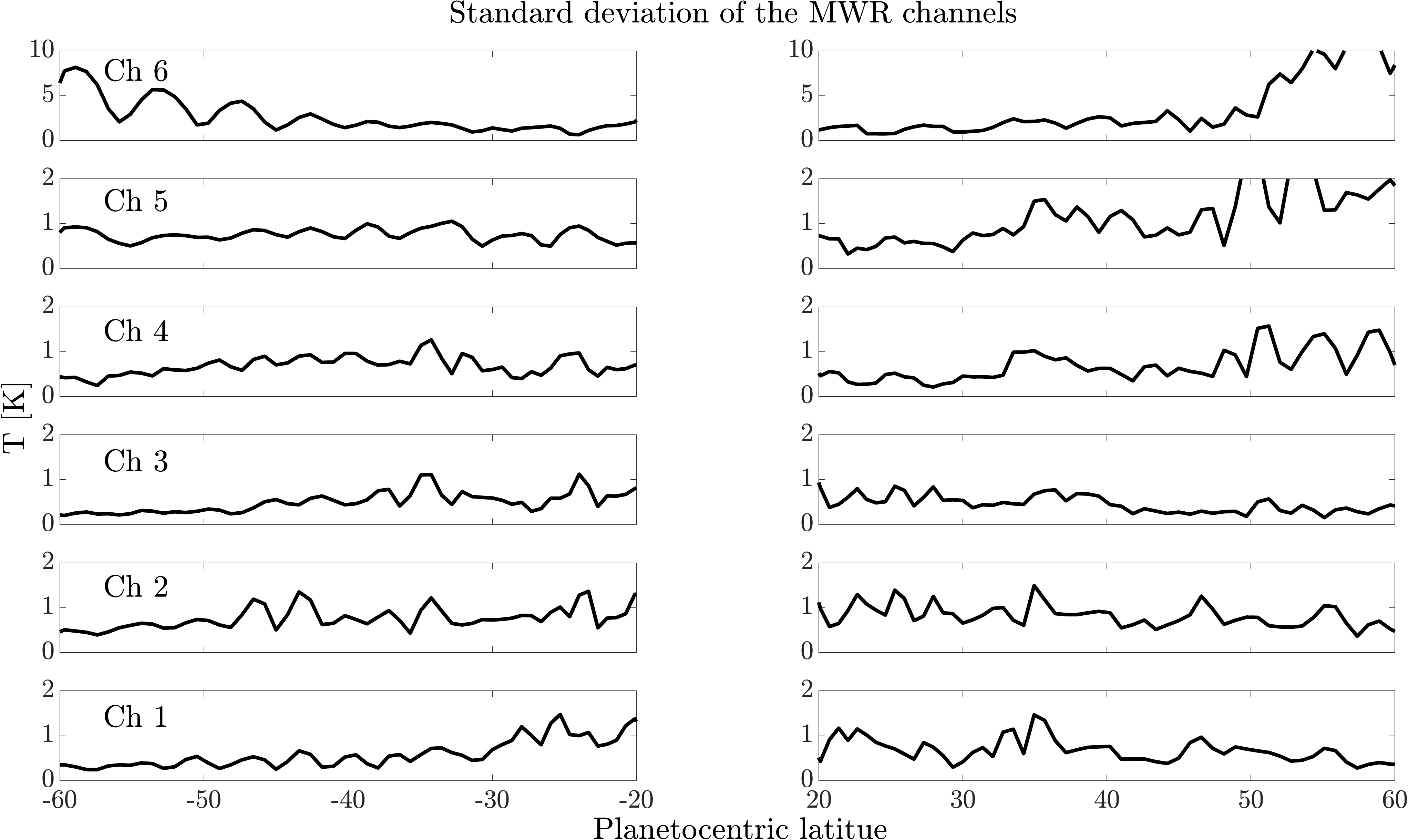}
\par\end{centering}
\caption{\label{fig:Tb std} The standard deviation (STD) of the nadir $T_{{\rm b}}$
in the midlatitudes computed from nine different perijoves through
PJs 1-12 \citep{oyafuso2020}, in the six MWR channels. The STD is
computed after the trend removal for each perijove, as detailed in
eq.~\ref{eq:Tb freq filter}. For channels 1-5, the STD values are
smaller than the variation seen in the data (Fig.~1h). Note that
although the STD values for channel 6 are higher, at some latitudes,
than the mean latitudinal variation (of the same channel), this channel
senses at altitudes which are above the cells identified in this study.}
\end{figure}
\begin{figure}
\begin{centering}
\includegraphics[width=0.7\textwidth]{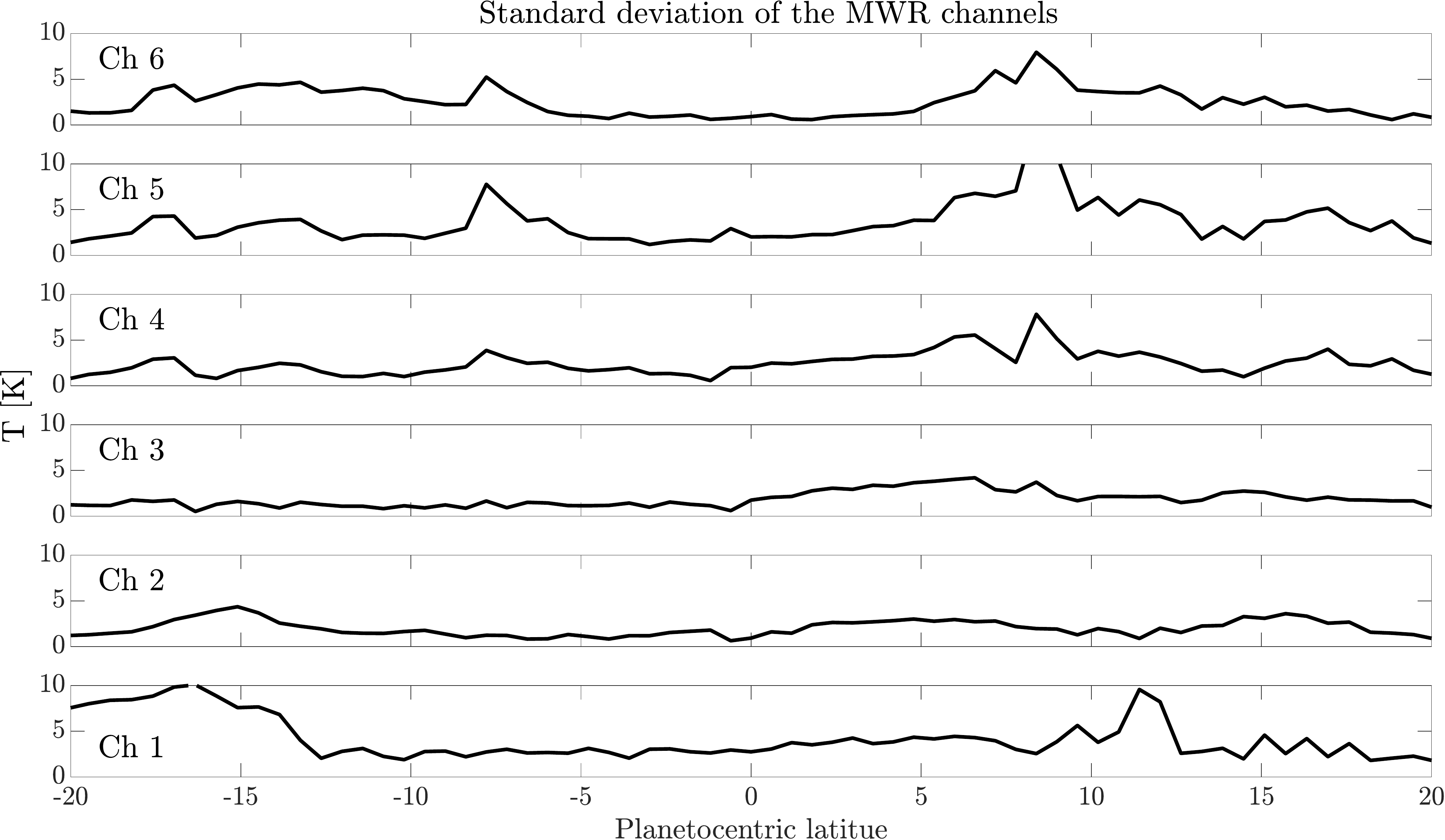}
\par\end{centering}
\caption{\label{fig:Tb std equator} The equatorial standard deviation of the
nadir $T_{{\rm b}}$ computed from nine different perijoves through
PJs 1-12 \citep{oyafuso2020}, in the six MWR channels. The STD is
computed on the original nadir coefficients.}
\end{figure}

As ammonia estimates by Juno's MWR for the high midlatitudes are not
yet available, we express the $T_{{\rm b}}$ measurements as ammonia
in order to examine the model results. For that, we define a reconstructed
ammonia distribution from Juno MWR data ($m_{a}^{({\rm data})}$),
used as a benchmark for the advection-relaxation model, constructed
by the mean ammonia calculated from MWR measurements of PJ1 ($M_{{\rm a}}$,
Fig.~2a) \citep{Li2017} and $T_{{\rm b}}$ measurements averaged
over multiple Juno orbits (PJs 1-12) \citep{oyafuso2020}. The standard
deviation between the perijoves is computed as a function of latitude
to validate that the latitudinal variations appearing in the $T_{{\rm b}}$
data are physical (Fig.~\ref{fig:Tb std} for the midlatitudes and
Fig.~\ref{fig:Tb std equator} for the equatorial region). We estimate
the $T_{{\rm b}}$ anomalies ($T_{{\rm b}}^{\prime}$) by removing
the cross-channel average ($T_{{\rm b},{\rm mean}}({\rm ch})$) from
each MWR channel, and then decompose $T_{{\rm b}}^{\prime}$ into
Legendre polynomials and reconstruct the anomalies without the low
polynomials to remove large scale variations. These variations, representing
equator to pole radiation differences, are not related to ammonia
variations by meridional cells \citep{oyafuso2020}. The reconstructed
ammonia is then

\begin{equation}
T_{{\rm b}}^{\prime}(\vartheta,{\rm ch})=T_{{\rm b}}(\vartheta,{\rm ch})-T_{{\rm b},{\rm mean}}({\rm ch})\cong\sum_{i=1}^{N}A_{i}(\vartheta,{\rm ch})P_{i}\left(\sin\vartheta\right),
\end{equation}

\begin{equation}
T_{{\rm b,rec}}^{\prime}(\vartheta,{\rm ch})=\sum_{i=30}^{N}A_{i}(\vartheta,{\rm ch})P_{i}\left(\sin\vartheta\right),\label{eq:Tb freq filter}
\end{equation}

\begin{equation}
m_{a}^{({\rm data})}(\vartheta,{\rm ch})=M_{{\rm a}}({\rm ch})-K({\rm ch})T_{{\rm b,rec}}^{\prime}(\vartheta,{\rm ch}),\label{eq:measured_ma}
\end{equation}
where $P_{i}$ are the Legendre polynomials, $A_{i}$ are the associated
coefficients, $N=200$ is the number of polynomials used and $K$
$[{\rm ppm}\cdot{\rm degrees}^{-1}]$ is a depth-dependent 'key',
optimized at each depth using Matlab's \texttt{'fmincon'} to best
fit the estimated ammonia distribution from PJ1 \citep{Li2017}. Note
that $T_{{\rm b}}$ is available between latitudes $90^{\circ}{\rm S}$
to $90^{\circ}{\rm N}$, therefore $P_{i=30}$ is equivalent to a
wavelength of approximately $12^{\circ}$ latitude. As the jets widths
are not larger the $8^{\circ}$ in latitude, this truncation removes
variations that are not due to the existence of meridional Ferrel-like
cells. The overall structure of $m_{a}^{({\rm data})}$ is very similar
to the ammonia map from PJ1 \citep{Li2017}, while the meridional
anomalies now represent well PJs 1-12 (Fig.~\ref{fig:pseudo_NH3}).
Finally, the resulting $K$ ranges between $5$ at $1$~bar to $0$
at depth and is used to estimate $m_{a}^{({\rm data})}$ at latitudes
$60^{\circ}{\rm S}$ to $60^{\circ}{\rm N}$. The profile is interpolated
between $1$ and $240$~bar, according to the relevant pressure levels
of each channel (Fig.~4d). These levels are estimated according to
the peak of the contribution function of each MWR channel, to give
that channels $\left\{ {\rm 1,2,3,4,5,6}\right\} $ correspond to
pressure levels of $\left\{ 240,30,9,3,1.5,0.7\right\} $~bar \citep{janssen2017,Bolton2017}.


\subsubsection{Cells construction and parameterization}

\begin{figure}
\begin{centering}
\includegraphics[width=0.8\textwidth]{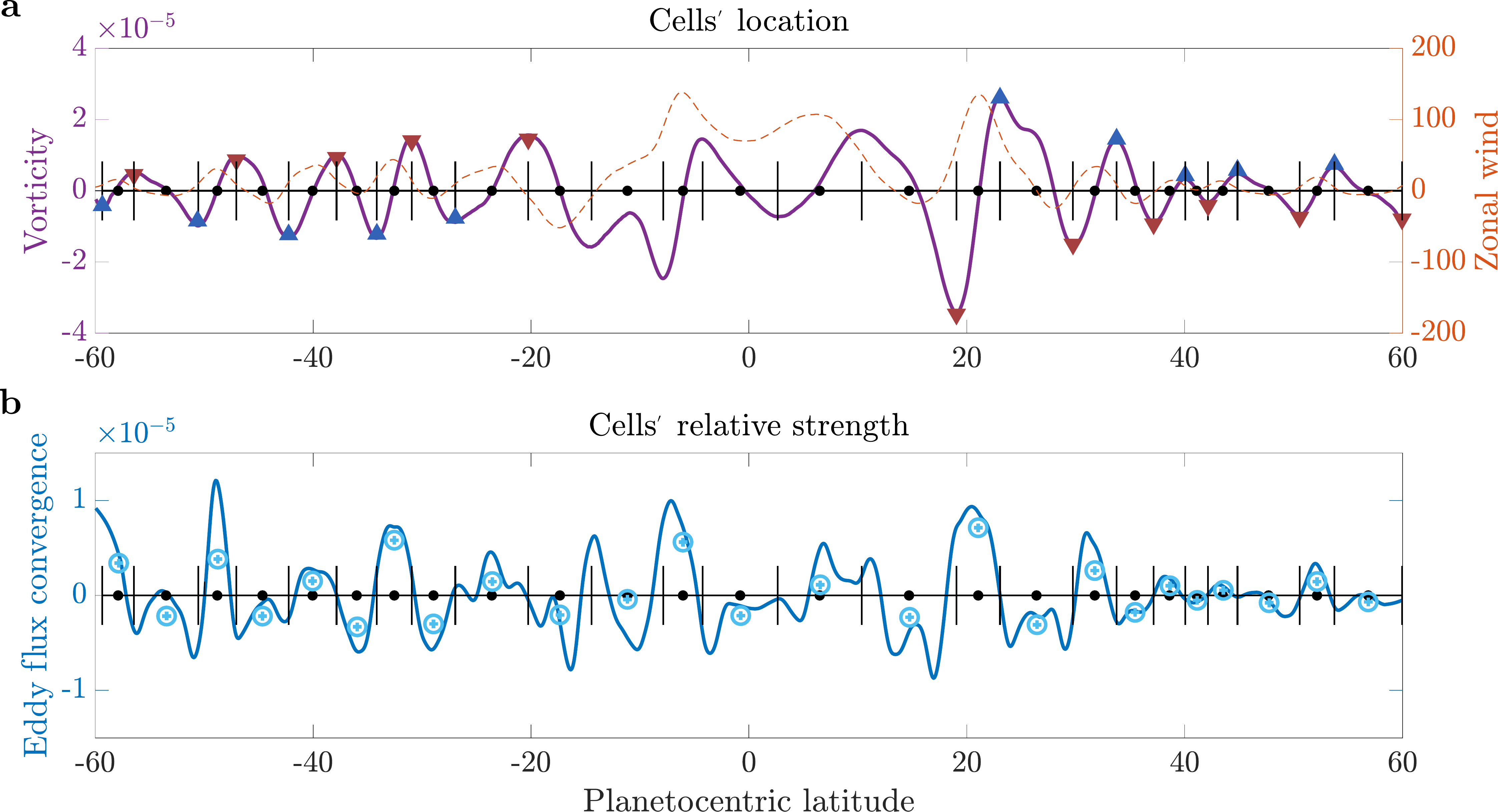}
\par\end{centering}
\caption{\label{fig:cells} (a) Vorticity (purple) and zonal wind (dashed,
orange). Blue (red) triangles represent vorticity peaks $\left(\nicefrac{\partial\zeta}{\partial y}=0\right)$,
where the vertical branches of the cells drive upward (downward) motion.
The cells' centers are positioned between vorticity peaks (black dots),
and the cells' extents ($2b_{k}$) are the distances between pairs
of black lines. (b) Eddy momentum flux convergence (blue) and the
cells' location as in panel a. Cells' relative strength is set by
the averaged value of eddy flux convergence within the cell (light
blue circles).}
\end{figure}

To describe the meridional cells in the simplest manner, we parameterize
each cell (indexed $k$) with an ellipse, according to a parameter
$l_{k}$ as follows 
\begin{equation}
l_{k}=\sqrt{\frac{\left(d-a\right)^{2}}{a^{2}}+\frac{\left(\vartheta-\vartheta_{k}\right)^{2}}{b_{k}^{2}}}.\label{eq:ellipse}
\end{equation}
Here, $d$ is defined as downward distance from the cloud level, and
$a$ and $b_{k}$ are the vertical and meridional extents of the cell,
respectively. $\vartheta_{k}$ is the latitude of the center of cell
$k$ (black dots in Fig.\ \ref{fig:cells}a,b). $b_{k}$ is set according
to half the width of cell $k$ (the distance between a black line
and a black dot in Fig.\ \ref{fig:cells}a,b). The outline of cell
$k$ (representing the path of the peak tangential velocity along
the cell) is thereby defined by $l_{k}=1$. For simplicity, the velocities
in a cell are defined using a normal distribution according to 
\begin{equation}
\begin{aligned}v_{k}= & V_{k}\exp\left[-\frac{1}{2}\left(\frac{l_{k}-1}{\widetilde{\sigma}}\right)^{2}\right]\sin\phi,\\
w_{k}= & V_{k}\exp\left[-\frac{1}{2}\left(\frac{l_{k}-1}{\widetilde{\sigma}}\right)^{2}\right]\cos\phi,\\
\widetilde{\sigma}= & \sigma\left(\cos^{2}\vartheta+\frac{b_{k}}{a}\sin^{2}\vartheta\right),
\end{aligned}
\end{equation}
where $\phi=\arctan\left[\frac{d-a}{R_{{\rm J}}\sin\left(\vartheta-\vartheta_{k}\right)}\right]$,
$R_{{\rm J}}$ is Jupiter's radius and $\sigma$ is a parameter for
the broadness of a cell's branch. $V_{k}$ represents the relative
strength (velocity) of cell $k$, parameterized according to the square
root of the averaged eddy momentum flux convergence within the cell
(Fig.\ \ref{fig:cells}b), and its sign represents the cell's direction
(clockwise/counter-clockwise), set according to the zonal wind sign
at the center of the cell (Fig.\ \ref{fig:cells}a).

\subsubsection{Optimization and numerical solution}

To solve for $m_{{\rm a}}$, Eq.\ 3 is discretized using finite differences
as 
\begin{equation}
\bar{w}_{i,j}\frac{m_{i-1,j}-m_{i+1,j}}{2dr}+\bar{v}_{i,j}\frac{m_{i,j+1}-m_{i,j-1}}{2R_{{\rm J}}d\vartheta}=-G_{i}\left(m_{i,j}-M_{i}\right),\label{eq: discrete conservation of mass}
\end{equation}
where the ``a'' subscript of $m$ and $M$ was removed for clarity.
Here $i$, and $j$ are indices for the grid points in the r and $\vartheta$
directions, respectively. $dr$ and $d\vartheta$ are the distances
between adjacent points in each direction. Eq.\ \ref{eq: discrete conservation of mass}
constitutes one of $n^{2}$ equations for $n^{2}$ variables, where
$n$ is the resolution of the grid in each direction. Eq.\ \ref{eq: discrete conservation of mass}
is rearranged in a matrix form as $Ax=b$ such that $m_{{\rm a}}$
can be calculated from $A^{-1}b$.

\begin{figure}[t]
\begin{centering}
\includegraphics[width=0.8\textwidth]{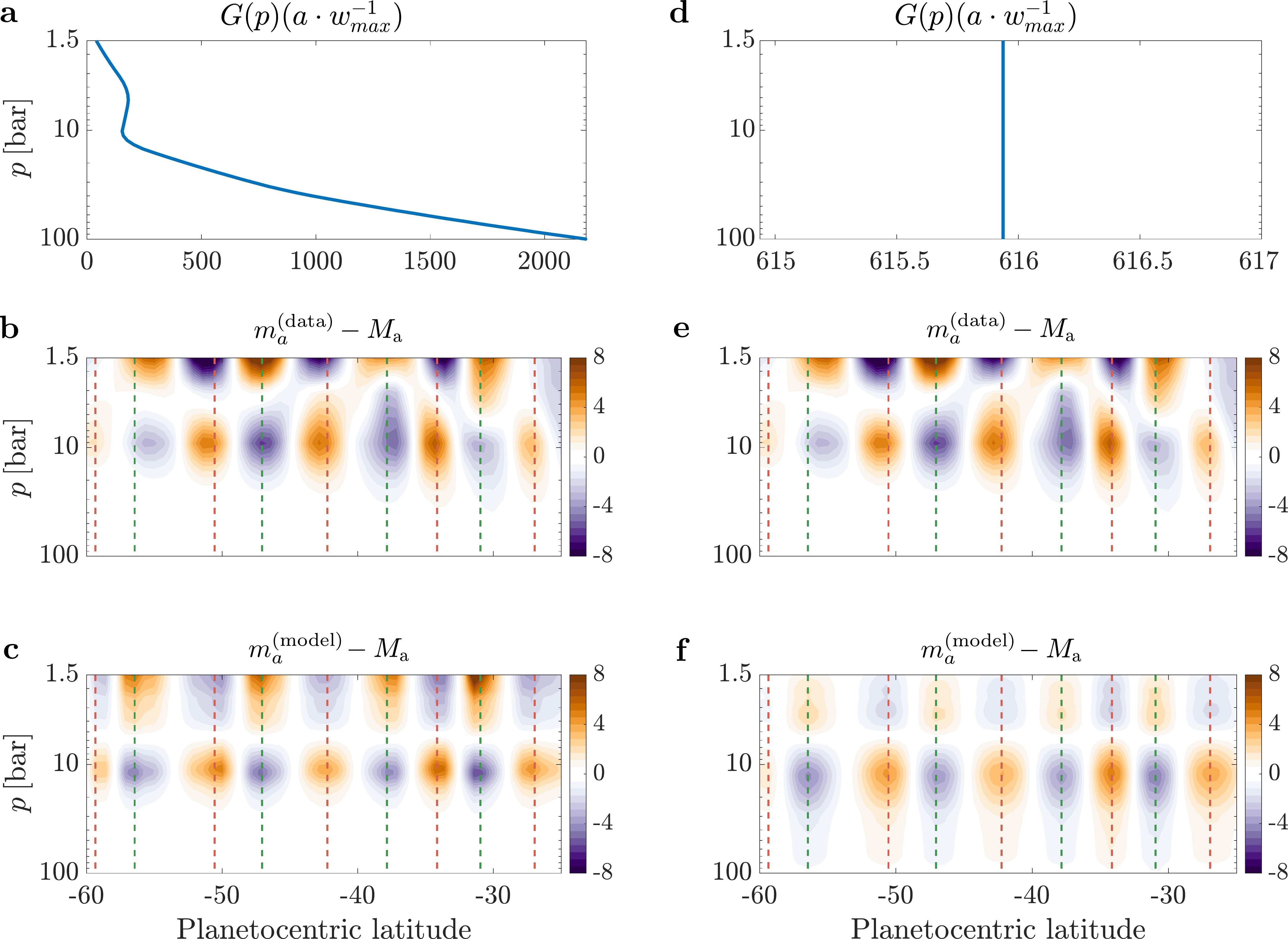}
\par\end{centering}
\caption{\label{fig:const_G} Comparison between model results with and without
vertical variation in $G.$\textbf{ }(a) The vertical variation of
the normalized source term $G$ used for Fig.~4c. b and e, $m_{a}^{({\rm data})}$
anomalies {[}ppm{]} in the SH. (c) The ammonia anomalies {[}ppm{]}
map produced by the advection-relaxation model with the source term
from panel a. (d)\textbf{ }Constant normalized source term $G$ in
Eq.~3. (f) The ammonia anomalies {[}ppm{]} map produced by the advection-relaxation
model with the source tern from panel d. In panels b, c, e and f the
vertical mean profile $M_{{\rm a}}$ is removed from the ammonia map
$m_{{\rm a}}$ and dashed red and green lines are the upward and downward
branches of the cells, respectively. The comparison reveals that although
the solution with varying source term (a) results in a better model
solution (\textbf{c}) compared to the measurements (b and d), the
essence of the anomalies (f) is well captured with a constant source
term (d).}
\end{figure}

The parameters $G_{i}$, $\sigma$ and $a$ are unknowns. For this,
the Matlab optimization function \texttt{'fmincon'} is used for deciding
$G_{i}$, $\sigma$ and $a$ to best reproduce the $m_{a}^{({\rm data})}$
map. The cost function 
\begin{equation}
f(G,\sigma,a)=\underset{i,j}{\sum}\left(\left|m_{i,j}^{({\rm model})}-m_{i,j}^{({\rm data})}\right|\right)^{2}
\end{equation}
is the measure used to find the optimal parameters. The resulting
value of $G$ is shown in Fig.~\ref{fig:const_G} for the case of
$G$ that is varying with depth, and for comparison, the solution
with a constant $G$ is shown as well. The value of $\sigma$ was
found to be $\sim0.85$ in both cases. The parameter $a$ was found
to be $\sim1600$ bar.

\subsubsection{Robustness analysis}

\begin{figure}[t]
\begin{centering}
\includegraphics[width=0.8\textwidth]{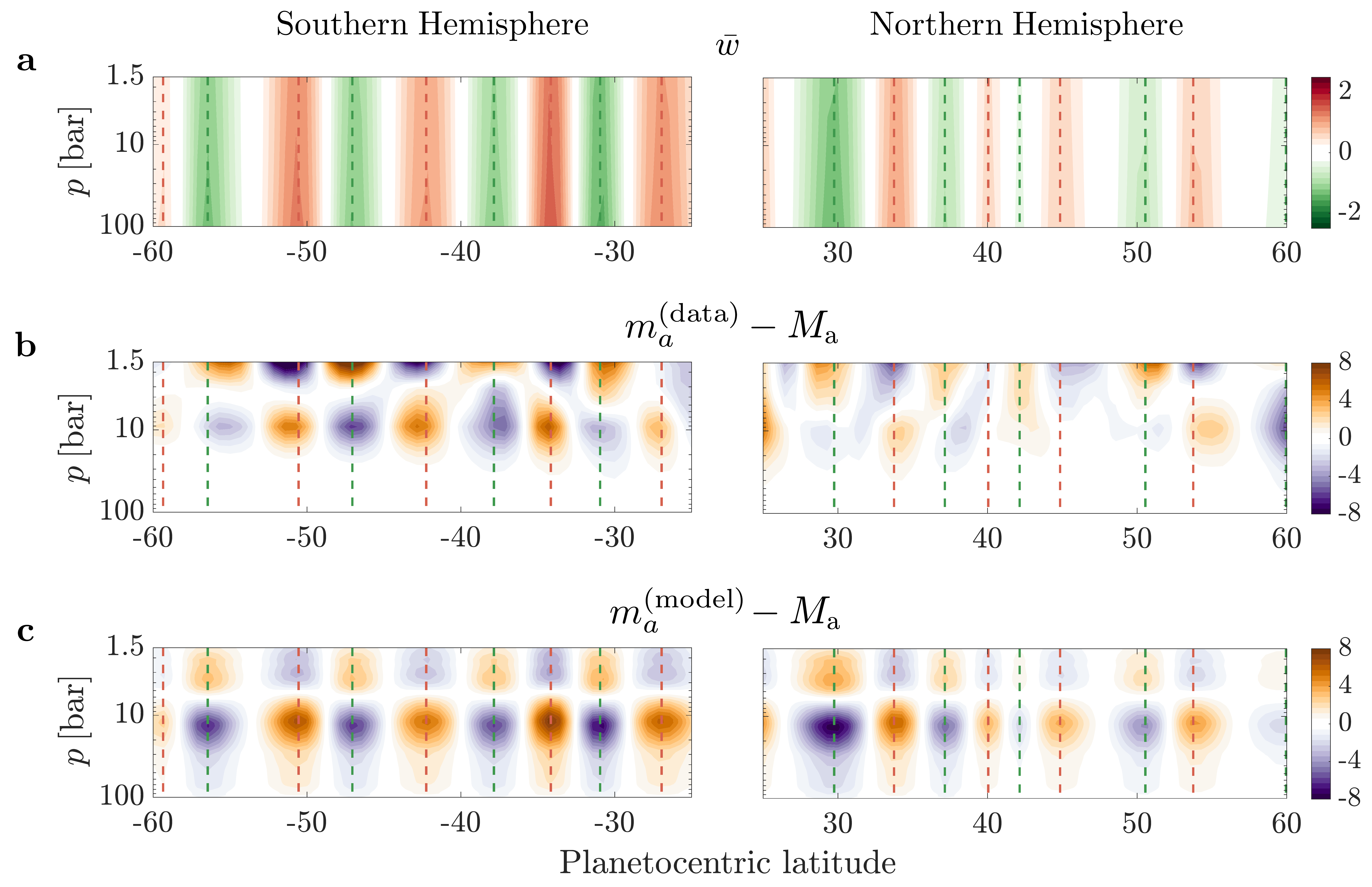}
\par\end{centering}
\caption{\label{fig:Model_no_opt}Model robustness analysis. An example for
a model run without optimization. The optimized variables in this
run are set manually according to physical considerations. The depth
of the cells ($a$) is set to $1,500$ km such that the cells extend
$3,000$ km in accordance with gravity measurements for the depth
of the zonal jets \citep{Kaspi2018}. The relaxation constant $G$
is set (without height dependence) from scaling argument by equating
the relaxation term in equation 3 to the vertical advection term,
leading to $G={\rm max}(w)/a$. The value of $\sigma$ is set to $0.5$.
(a) The normalized vertical velocity as a function of latitude and
pressure. (b) The ammonia anomalies map that was reconstructed from
Juno's MWR measurements. (c) The map of ammonia anomalies resultant
from the degenerated model. It can be seen that the optimization procedure
doesn't change the nature of the results which is robust. The structure
of the anomalies stays largely the same both in this figure and in
Fig.~4, and it stems mostly from the latitudinal structure of the
wind and the vertical stratification of the ammonia, both being derived
from observations.}
\end{figure}
To validate that the model results are robust and not sensitive to
the specific parameters found by the optimization analysis, $m_{i,j}^{({\rm model})}$
was also solved from equation 3 using a chosen set of parameters instead
of the optimized set. The depth of the cells was chosen to be $3000$~km
($a=1500\,{\rm km}$), in accordance with the depth of the jets that
was estimated from gravity measurements \citep{Kaspi2018}. The width
of the cells was chosen such that one standard deviation covers half
a cell's width ($\sigma=0.5$). $G$ is degenerated into a constant
(no dependence in $z$) and is set, from a scaling argument, as $G={\rm max}(w)/a\,{\rm [s^{-1}]}$.
As seen in the results (Fig.\ \ref{fig:Model_no_opt}), the modeled
ammonia anomalies map still predicts the data convincingly. The main
difference is the depth where the sign of the ammonia anomalies flip
(between $3$ and $6$ bar), which is now only controlled by the input
$M_{{\rm a}}$ \citep{Li2017}, and could not be 'corrected' by a
depth-dependent relaxation time scale. It can be seen that the essence
of the circulation cells is still very apparent in the results.

\section{Equatorial region analysis}

\subsection{Estimate for the extent of the equatorial region}

The tangent cylinder is the projection (along the axis of rotation)
of the planet\textquoteright s solid-body rotating core on the outer
shell. The equatorial latitudes lie outside of the tangent cylinder,
and is thereby characterized by a different dynamical regime than
that of the midlatitudes. To separate quantitatively the midlatitudes,
positioned within the tangent cylinder, from the equatorial region,
it is required to know the depth of the atmosphere ($D_{{\rm atm}}$)
and the radius of the planet ($R$). The latitudes of the cylinder's
edge ($\alpha$) can then be derived from geometrical considerations
as

\begin{equation}
\alpha=\arccos\left(1-\frac{D_{{\rm atm}}}{R}\right).\label{eq:tangent_cylinder}
\end{equation}
Gravity analysis reveals that Jupiter's atmosphere is approximately
$3000\,\,{\rm km}$ deep \citep{Guillot2018,Kaspi2018}. Substituting
$D_{{\rm atm}}=3000\,\,{\rm km}$ and $R=R_{{\rm J}}=70,000\,\,{\rm km}$
in Eq.~\ref{eq:tangent_cylinder} gives $\alpha=\pm16.8^{\circ}$.
This means that fluid columns parallel to the axis of rotation in
the latitude range $-17^{\circ}\leq\vartheta\leq17^{\circ}$, can
theoretically extend uninterruptedly between the hemispheres.

\subsection{Theory for the leading balance in the Jovian equatorial region}

Starting from the primitive equations\citep{Vallis2006}, the continuity
and zonal momentum equations in spherical coordinates are

\begin{equation}
\frac{\partial\rho}{\partial t}+\rho\left(\frac{1}{R_{{\rm J}}\cos\vartheta}\frac{\partial u}{\partial\lambda}+\frac{1}{R_{{\rm J}}\cos\vartheta}\frac{\partial}{\partial\vartheta}\left(v\cos\vartheta\right)+\frac{\partial w}{\partial r}\right)+\frac{u}{R_{{\rm J}}\cos\vartheta}\frac{\partial\rho}{\partial\lambda}+\frac{v}{R_{{\rm J}}}\frac{\partial\rho}{\partial\vartheta}+w\frac{\partial\rho}{\partial r}=0,\label{eq:Coninuity-primitive}
\end{equation}
 and
\begin{equation}
\frac{\partial u}{\partial t}+\frac{u}{R_{{\rm J}}\cos\vartheta}\frac{\partial u}{\partial\lambda}+\frac{v}{R_{{\rm J}}}\frac{\partial u}{\partial\vartheta}+w\frac{\partial u}{\partial r}-2\Omega\left(\sin\vartheta\,v-\cos\vartheta\,w\right)-\frac{uv}{R_{{\rm J}}}\tan\vartheta=-\frac{1}{R_{{\rm J}}\rho\cos\vartheta}\frac{\partial p}{\partial\lambda},\label{eq: Zonal momentum - primitive}
\end{equation}
respectively, where $\lambda$ is longitude. The mean density ($\rho_{m}$)
is assumed to only change with $r$, and anomalies from it are assumed
to be much smaller than the mean value. In addition, the meridional
derivatives and the meridional velocity are assumed very small near
the equator, relative to the other terms. This assumption is based
on the symmetry around the equator. Assuming also a steady state,
Eq.~\ref{eq:Coninuity-primitive} and Eq.~\ref{eq: Zonal momentum - primitive},
evaluated on the equatorial plane ($\vartheta=0^{\circ}$), are then

\begin{equation}
\rho_{m}\left(\frac{1}{R_{{\rm J}}}\frac{\partial u}{\partial\lambda}+\frac{\partial w}{\partial r}\right)+w\frac{\partial\rho_{m}}{\partial r}=0,\label{eq:Cont_zonal average}
\end{equation}
and
\begin{equation}
\frac{u}{R_{{\rm J}}}\frac{\partial u}{\partial\lambda}+w\frac{\partial u}{\partial r}+2\Omega w=-\frac{1}{R_{{\rm J}}\rho_{m}}\frac{\partial p}{\partial\lambda}.\label{eq:zonal momentum _zonal average}
\end{equation}
Eq.~\ref{eq:zonal momentum _zonal average} can equivalently be represented
as 
\begin{equation}
\frac{1}{R_{{\rm J}}}\frac{\partial u^{2}}{\partial\lambda}+\frac{1}{\rho_{m}}\frac{\partial\left(wu\rho_{m}\right)}{\partial r}-\frac{u}{\rho_{m}}\left(\rho_{m}\frac{1}{R_{{\rm J}}}\frac{\partial u}{\partial\lambda}+\rho_{m}\frac{\partial w}{\partial r}+w\frac{\partial\rho_{m}}{\partial r}\right)+2\Omega w=-\frac{1}{R_{{\rm J}}\rho_{m}}\frac{\partial p}{\partial\lambda},\label{eq:zonal momentum _zonal average-2}
\end{equation}
where the third term vanishes according to Eq.~\ref{eq:Cont_zonal average}.
Next, the velocities are decomposed via Reynolds decomposition ($u=u'+\overline{u},\,\,\,\,w=w'+\overline{w}$),
such that the zonal mean of Eq.~\ref{eq:zonal momentum _zonal average-2}
gives
\begin{equation}
-\frac{\partial\left(\overline{w'u'}\rho_{m}\right)}{\partial r}=\overline{w}\frac{\partial\left(\overline{u}\rho_{m}\right)}{\partial r}+2\rho_{m}\Omega\overline{w}+\rho_{m}\overline{u}\frac{\partial\overline{w}}{\partial r}.\label{eq:Zonal momentum_3}
\end{equation}
This shows that the momentum originating from the eddy momentum flux
convergence $\left(\frac{\partial\left(\overline{w'u'}\rho_{m}\right)}{\partial r}<0\right)$,
which drives the equatorial superrotation \citep{Kaspi2009}, is divided
between the growing equatorial super-rotating jet $\left(\frac{\partial\left(\overline{u}\rho_{m}\right)}{\partial r}>0,\,\overline{u}>0\right)$,
the Coriolis force and another residual term. The growing equatorial
jet has been shown in many numerical simulations of superrotation
\citep{Heimpel2005,Kaspi2009,Gastine2014}. It is a good assumption
that each of the terms on the right side is of smaller magnitude than
the source term on the left side. Finally, rearranging Eq.~\ref{eq:Zonal momentum_3}
gives
\begin{equation}
\overline{w}=-\frac{1}{\underset{>0}{\underbrace{\frac{\partial\left(\overline{u}\rho_{m}\right)}{\partial r}}}+\underset{>0}{\underbrace{2\rho_{m}\Omega}}}\underset{<0}{\underbrace{\left(\frac{\partial\left(\overline{w'u'}\rho_{m}\right)}{\partial r}+\rho_{m}\overline{u}\frac{\partial\overline{w}}{\partial r}\right)}}>0.\label{eq: sign of wm}
\end{equation}
To further simplify, Eq.~\ref{eq: sign of wm} can be shown for the
case of small Rossby number:
\begin{equation}
\overline{w}=-\frac{1}{2\rho_{m}\Omega}\frac{\partial\left(\overline{w'u'}\rho_{m}\right)}{\partial r}>0.\label{eq: sign of wm-1}
\end{equation}
This implies that the mean upwelling ($\overline{w}$) correlates
with eddy momentum flux convergence, and therefore with the equatorial
superrotating jet at the equatorial region. Since the superrotating
jet is supposed to be driven by angular momentum fluxes in the direction
perpendicular to the rotation axis \citep{Heimpel2005,Kaspi2009,Schneider2009},
converging in the equatorial region, Eq.~\ref{eq: sign of wm-1}
would take the more general form 
\begin{equation}
\overline{w_{\perp}}=-\frac{1}{2\rho_{m}\Omega}\partial_{\perp}\left(\overline{w_{\perp}'u'}\rho_{m}\right)>0,\label{eq: sign of wm perp}
\end{equation}
where $w_{\perp}$ and $\partial_{\perp}$ are the velocity and the
gradient in the direction perpendicular to the axis of rotation, i.e.,
$w_{\perp}=w\cos\vartheta+v\sin\vartheta$ and $\partial_{\perp}=\left(\cos\vartheta\right)\partial_{r}+\left(r^{-1}\sin\vartheta\right)\partial_{\vartheta}$.

\begin{figure}
\begin{centering}
\includegraphics[width=0.35\textwidth]{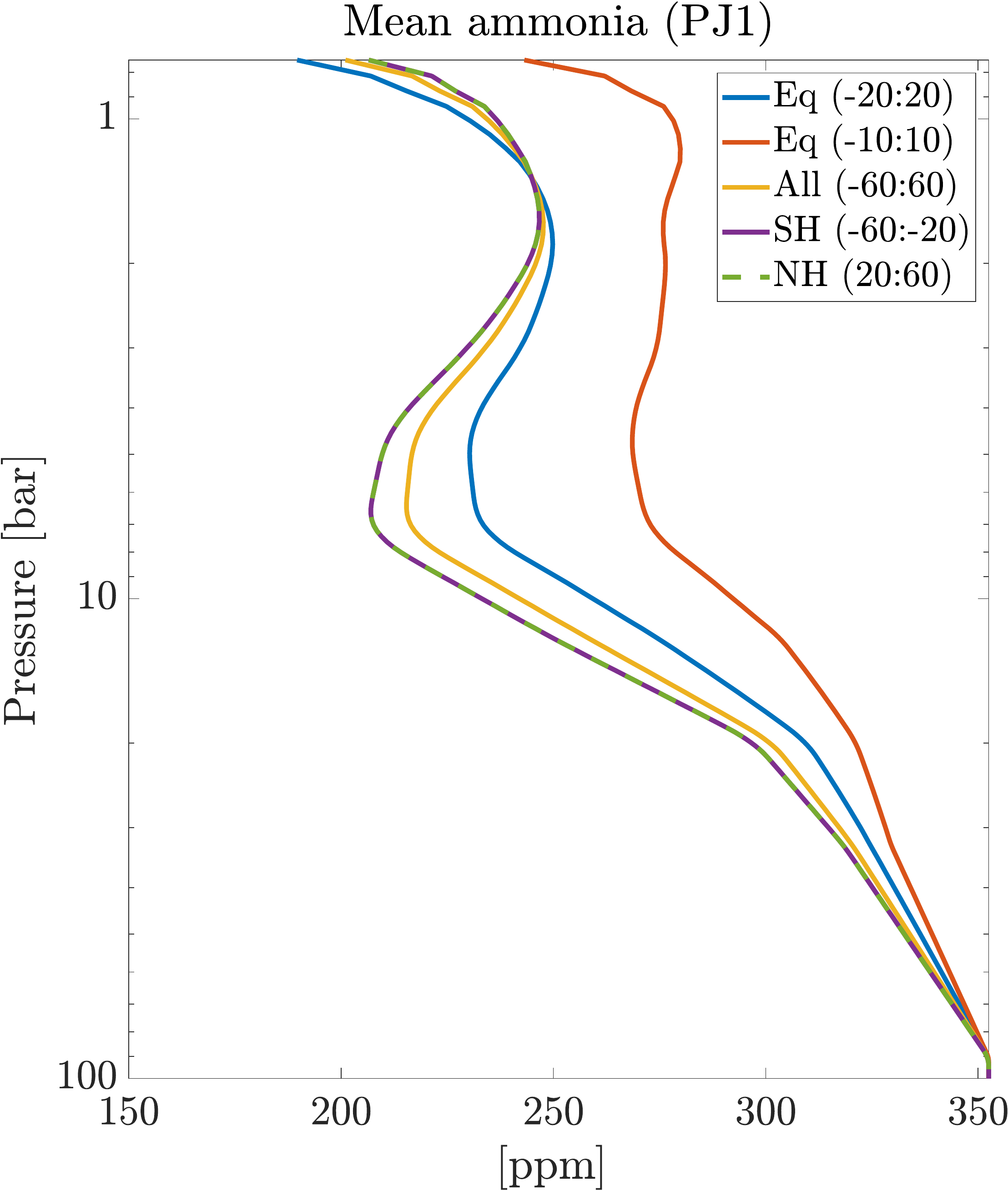}
\par\end{centering}
\caption{\label{fig:NH3 bar} Meridional averaged ammonia values at different
latitudinal regions. The lines are calculated according to the inferred
ammonia map from PJ1 \citep{Li2017}. Each line is averaged at the
latitudinal range described in the legend.}
\end{figure}

\end{document}